\documentclass[aps,prc,twocolumn,floatfix,nofootinbib,superscriptaddress]{revtex4-1}  
\usepackage{amsmath}
\usepackage{amsfonts}
\usepackage{amssymb}
\usepackage{graphicx}
\usepackage{epstopdf}
\usepackage{xcolor}
\usepackage{appendix}
{\LARGE }
\usepackage{hyperref}
\hypersetup{
	colorlinks=true,
	linkcolor=blue,
	filecolor=blue,
	urlcolor=blue,
	citecolor=blue,
    pdfstartview=FitH,
    pdfpagemode=UseNone,
}

\begin{document}

\title{ \bf{ Asymptotic normalization coefficients  for $\alpha+ {}^{12}{\rm C}$ synthesis  and the $S$-factor for ${}^{12}{\rm C}(\alpha, \,\gamma){}^{16}{\rm O}$  radiative capture}}
\author{A. M. Mukhamedzhanov}
\email{akram@comp.tamu.edu}
\affiliation{Cyclotron Institute, Texas A$\&$M University, College Station, TX 77843, USA}
\author{R. J. deBoer}
\affiliation{Department of Physics and Astronomy and the Joint Institute for Nuclear Astrophysics, University of Notre Dame, Notre Dame, Indiana 46556, USA}
\author {B. F. Irgaziev}
\affiliation{Theoretical Physics Department, National University of Uzbekistan, Tashkent 100174, Uzbekistan}
\author{L. D. Blokhintsev}
\affiliation{Skobeltsyn Institute of Nuclear Physics, Lomonosov Moscow State University,  Moscow 119991, Russia}
\author{A. S. Kadyrov}
\affiliation{Department of Physics and Astronomy, Curtin University, GPO Box U1987, Perth, WA 6845, Australia}
\affiliation{Institute of Nuclear Physics, 
Ulugbek, Tashkent 100214, Uzbekistan}
\author{D. A. Savin}
\affiliation{Skobeltsyn Institute of Nuclear Physics, Lomonosov Moscow State University,  Moscow 119991, Russia}

\begin{abstract}
{\bf{Background}:} The  $^{12}{\rm C}(\alpha,\gamma)^{16}$O  reaction, determining the survival of carbon in red giants,  is of interest for nuclear reaction  theory and nuclear astrophysics. A specific feature of the $^{16}$O nuclear structure  is the presence of two subthreshold bound states,   (6.92 MeV, 2$^+$) and  (7.12 MeV, 1$^-$), that dominate the behavior of the low-energy $S$-factor. The strength of these subthreshold states  is determined by their asymptotic normalization coefficients (ANCs), which need to be known with high accuracy. \\
{\bf{Purpose}:}   The objective of this research is to examine how the subthreshold and ground-state ANCs impact the low-energy $S$-factor, especially at the key astrophysical energy of $300$ keV.\\
{\bf{Method:}}  
The $S$-factors are calculated within the framework of the $R$-matrix method  using the AZURE2 code.\\  
{\bf{Conclusion}:}  Our total $S$-factor takes into account the $E1$ and $E2$ transitions to  the ground state of $^{16}$O including the interference of the subthreshold and higher resonances, which also interfere with the corresponding direct captures, and cascade radiative captures to the ground state of $^{16}$O through four subthreshold states: $0_2^+,\,3^-,\, 2^+$ and $1^-$. To evaluate the impact of subthreshold ANCs on the low-energy $S$-factor, we employ two sets of the ANCs. The first selection, which offers higher ANC values, is attained through the extrapolation process [Blokhintsev {\it et al.}, Eur. Phys. J.  {\bf 59},  162 (2023)]. The set with low ANC values was employed by deBoer {\it et al.} [Rev. Mod. Phys. {\bf 89},  035007 (2017)]. A detailed comparison of the  $S$-factors at the most effective astrophysical energy of 300 keV is provided, along with an investigation into how the ground-state ANC affects this $S$-factor.
\end{abstract}

\maketitle


\section{Introduction}

The ${}^{12}{\rm C}/{}^{16}{\rm O}$  ratio in red giant stars has been attracting substantial scientific attention for a long time  \cite{RolfsRodney,deBoer}.  While ${}^{12}{\rm C}$  is formed via the triple-$\alpha$ fusion,  ${}^{16}{\rm O}$ is the result of the $\,{}^{12}{\rm C}(\alpha, \,\gamma){}^{16}{\rm O}$ radiative capture reaction,  which determines the survival of carbon.  Numerous attempts to obtain the astrophysical factor of
the  ${}^{12}{\rm C}(\alpha, \,\gamma){}^{16}{\rm O}$ reaction, both experimental and theoretical, have been made for almost 50 years 
(see \cite{RolfsRodney,deBoer, Kettner,redder,Kremer,BarkerKajino,Ouellet,azuma,Brune,kunz,assuncao,Balhout,plug,sayre,Schurmann,gai2013,Avila,Shen} and references therein).  The latest  comprehensive and thorough review 
of the state of the art  has been presented in \cite{deBoer}.

The main goal of the  research is to obtain the astrophysical $S$-factor   for the  ${}^{12}{\rm C}(\alpha, \,\gamma){}^{16}{\rm O}$   radiative capture  in the Gamow window with the  most effective astrophysical $\alpha-{}^{12}{\rm C}$ relative kinetic energy energy  $E= 300$ keV  with the accuracy  $\,  <10\%$.  However,  this goal is far from being achieved because at this energy  direct measurements   are hardly  feasible  due to the extremely   small cross  section\footnote{The lowest  energy at which the $S$-factor was measured  is  $\approx 1$ MeV. }.  Extrapolating down  the  experimental  data  available at energies $E >1$  MeV  to the low-energy  region  allows one to obtain  the $S(300\,{\rm keV})$-factor. The most popular method of the extrapolation is the $R$-matrix approach \cite{deBoer}. 
It provides a way to control contributions of different 
interfering
mechanisms of the radiative capture and  impact of different input parameters.

It is very well known (see, for example,  \cite{deBoer,Brune}  and references therein)  that one of the important parameters for  extrapolation  of the experimental data  to the low-energy region and determination  of the $S(300\,{\rm keV})$-factor  are  the asymptotic normalization coefficients  (ANCs)  for the  $\alpha+ {}^{12}{\rm C} \to  {}^{16}{\rm O}(7.12\, {\rm MeV}, 1^{-})$    and    $\alpha+ {}^{12}{\rm C} \to  {}^{16}{\rm O}(6.92\, {\rm MeV}, 2^{+})$   
 syntheses leading to the formation of two subthreshold (that is, near the  $\alpha + {}^{12}{\rm C}\,$ threshold)  bound  states, $1^{-}$ and $2^{+}$.  The binding  energies  of these  bound states are   $\varepsilon_{s1} = 0.045$ MeV   and   $\varepsilon_{s2} = 0.2449$ MeV,
 respectively. Since these subthreshold states are the closest levels to  the low-energy  region  $E <0.5$ MeV  (the nearest $1^{-}$ and $2^{+}$ resonances are located at  $2.43$  and  $2.68$  MeV, respectively),  they govern the behavior of the low-energy $S$-factor of the  ${}^{12}{\rm C}(\alpha, \,\gamma){}^{16}{\rm O}$ reaction (see \cite{deBoer}  and references therein)  through the dominating $\,E1$ and $E2\,$ resonance radiative captures  to the ground state of ${}^{16}{\rm O}.$
Hence  the uncertainties of the  subthreshold ANCs  should play an important  role  in the   determination of the  uncertainty  of the  total  
 $S(300\,{\rm keV})$-factor.  

There are the following important problems  regarding  the impact of the subthreshold  ANCs  on the low-energy $S$-factors. 
First, the ANCs available in the literature  vary quite significantly   (see  Table \ref{Table_ANCs}   below) and the question is how much the variation of the subthreshold ANCs affects  the low-energy $S$-factors.  The second problem is  the necessity to determine the contribution of the $1^{-}$  and $2^{+}$   subthreshod resonances (SRs),  which are controlled by the subthreshold  ANCs,  to the total  low-energy $S$-factor, and, in particular,  $S(300\,{\rm keV})$-factors.
  Answering this question  will allow us to understand  the  contribution of the uncertainties of the subrthreshold ANCs to the budget of the uncertainty  of the  low-energy $S$-factor, especially,  the  $S(300\,{\rm keV})$-factor. 
The third problem is to understand the effect of the interference  of  the  SRs  with  higher resonances  and direct capture  for the $E1$  and $E2$ transitions.  Finally,  we need to understand  the correlated effect of the subthreshold and ground-state ANCs on the low-energy $S$-factor and, in particular, on the $S(300\,{\rm keV})$-factor.

A novel method was developed and employed in a sequence of papers \cite{Blokh2017, Blokh2018, BKMS5, Blokh2023} to ascertain the ANCs through extrapolation of elastic scattering phase shifts to subthreshold bound-state poles. Specifically, in  Refs.  \cite{BKMS5,Blokh2023}, this technique was applied to determine the ANCs for $\alpha$-particle removal from the four subthreshold states of ${}^{16}{\rm O}$: $0_{2}^{+},\,3^{-},\, 2^{+}$ and $1^{-}$.
In this paper, we address all the aforementioned problems by performing comprehensive $R$-matrix  calculations.   We use the subthreshold ANCs  found by the extrapolation procedure  and the ones taken from Ref. \cite{deBoer},  together with two  ground-state 
ANCs of  ${}^{16}{\rm  O}$,  a low value of  $58$ fm${}^{-1/2}$   \cite{deBoer}  and  a  high value of  $337$ fm${}^{-1/2}$  \cite{Shen}.   
Unless stated otherwise, we adopt the unit system where $\hbar=c=1.\,$  
\section{Methodology}  
\subsection{Subthreshold  resonances}
\label{SRS1}
 
Nuclear excited states below the particle emission threshold (subthreshold states) typically undergo 
$\gamma$ decay to lower lying states.  Besides, the presence of the subthreshold states generate a new mechanism of the radiative capture to the low lying states in which the subthreshold states exhibit themselves as  SRs.  
Below we can give a simple explanation of the capture through the SR. 

Let us consider  two spinless, structereless particles in the continuum forming a  bound state by emitting a photon.  A very simplified, schematical equation for the  amplitude  corresponding to  the low-energy  radiative transition  to the ground state  is 
 \begin{align}
 M_{NR}  \sim  <\varphi_{f}\big| {\cal  H}_{el}^{L}   \big|\psi>.
 \label{potmod1}
 \end{align}
Here  $\psi$  is the scattering wave function,  $\,\varphi_{f}$ is  the ground bound-state wave function of the interacting particles. The electromagnetic interaction Hamiltonian $ {\cal H}_{el}^{L} $ is associated with the transition characterized by multipolarity $L$. 

There are two possible  approaches to calculate  $\,\psi(r)\,$. 
One of them is the two-body  potential model, which allows us to  simultaneously  treat nonresonant,
SR and resonance transitions in the given partial wave (see \cite{Dubovichenko} and the references therein). However,  the adopted two-body potential should reproduce simultaneously available experimental scattering phase shifts,  binding energies and the corresponding  ANCs, and the resonances in the given partial wave.   Besides, the two-body potential model may not be good enough to treat  the nuclear interior where the  many-body approach is required.  
Also accurate accounting for interference effects  continues to be a problem.

The second approach is the $R$-matrix formalism, which we utilize in this paper. 
We underscore the advantages  of the $R$-matrix method: \\
1. By separating  different mechanisms we can explicitly  single out the ANCs of the subthreshold bound state and the ground-state ANC as the normalization coefficients of the corresponding  amplitudes. \\
2. Another advantage is that the internal part of the  SR radiative amplitude is parametrized in terms of  the  internal  radiative width amplitude which can be calculated using the experimental radiative widths and the channel (external) counterparts or  can be a fitting parameter. \\
3. Finally, the $R$-matrix  method allows one to take into account  the interference effects in a straightforward manner.

\begin{figure}[tbp]
\includegraphics[width=5.0cm]{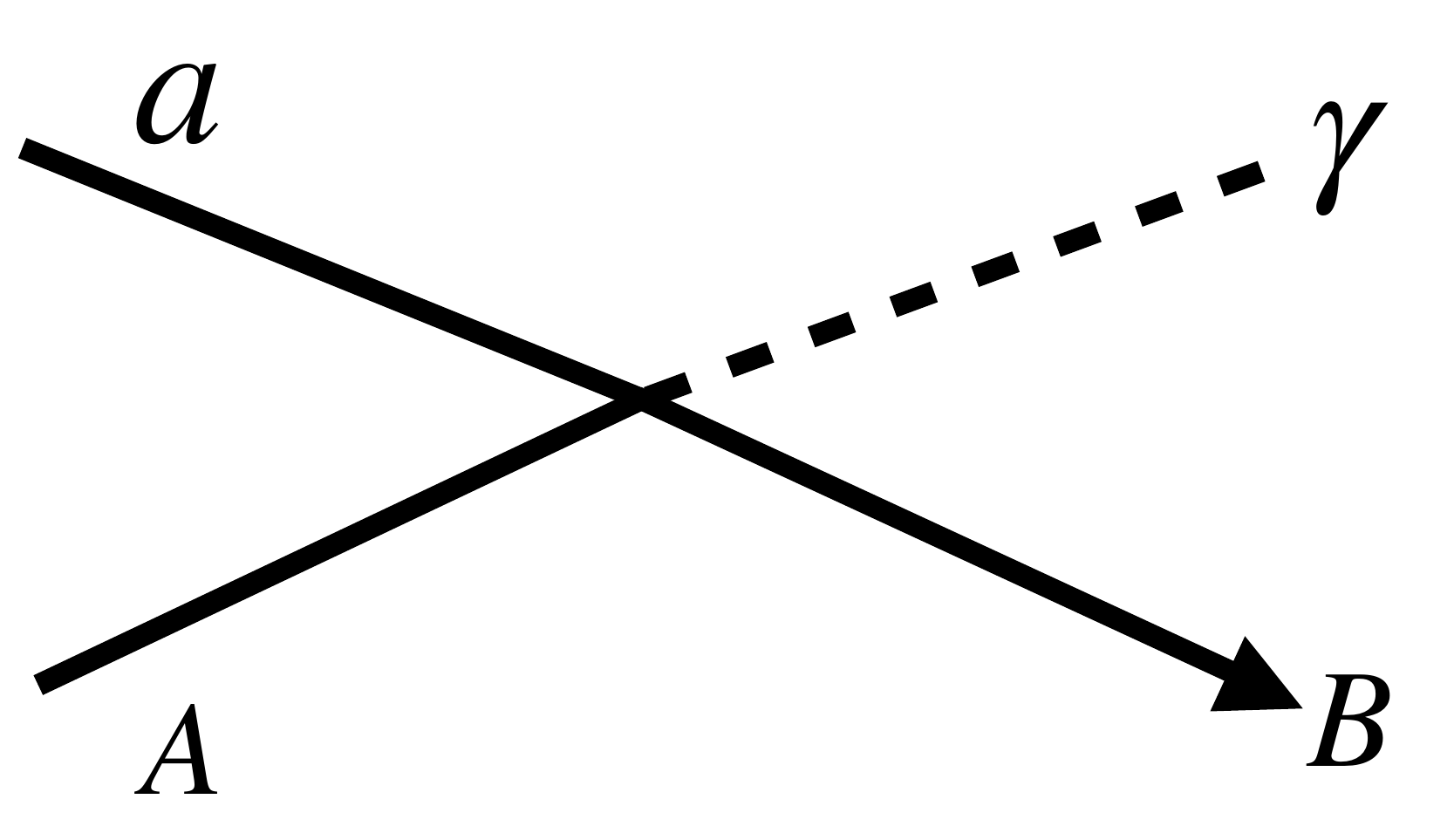}
\caption{The  diagram describing the nonresonant radiative capture ${}^{12}{\rm C}(\alpha,\,\gamma){}^{16}{\rm O}$. The short-dashed line represents the emitted photon.}
\label{fig_NR} 
\end{figure}

We rewrite Eq. (\ref{potmod1})  in a form convenient for the $R$-matrix formalism: 
 \begin{align}
  M_{NR}  \sim  <\varphi_{f}\big| {\cal  H}_{el}^{L}   \big| I - {\mathbb S}\,O> ,
 \label{scemMradcpt1}
 \end{align}
 where  $I$ and $O$  are the incident and outgoing  scattered waves in the initial channel,  $\,{\mathbb  S}\,$ is the elastic scattering $S$-matrix element.   The diagram shown in  Fig \ref{fig_NR}  describes the nonresonant
 radiative capture.

\begin{figure}[tbp]
\includegraphics[width=6.0 cm]{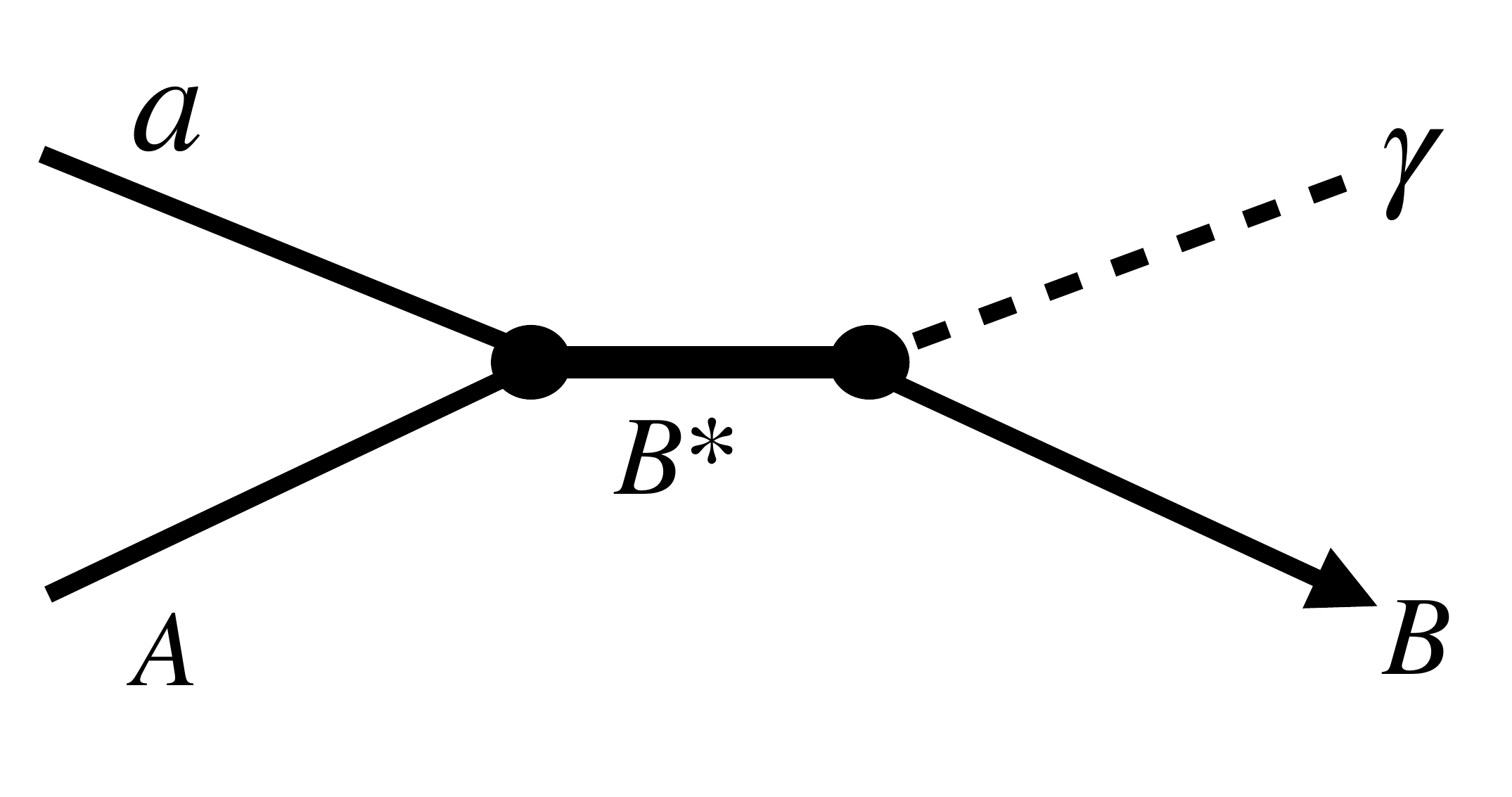}
\caption{The pole diagram describing the radiative capture  ${}^{12}{\rm C}(\alpha,\,\gamma){}^{16}{\rm O}$  through  the SR  $\,B^{*}$.  The thick black line is the resonance propagator,  two black spheres are the form factors and the short-dashed line is the emitted photon. }
\label{fig_pole} 
\end{figure}

From Eq.  (\ref{scemMradcpt1})  one can obtain the  radiative capture amplitude 
proceeding through the SR.   The presence of the subthreshold bound state  generates  a pole in the
 ${\mathbb  S}$-matrix element  at $E= -\varepsilon_{s},$ 
 where $\varepsilon_{s}$ is the binding energy of the subthreshold bound  state,
 affecting the second term of the initial scattering wave function in Eq. (\ref{scemMradcpt1}). 
The impact of the subthreshold bound state on the elastic scattering $S$-matrix behavior, controlled by a subthreshold pole, becomes more pronounced as the energy approaches zero:
 \begin{align}
 {\mathbb S} \stackrel{E \to +0}{\sim}  \frac{{\cal A}_{s}}{E+ \varepsilon_{s}   +  i\, \frac{\Gamma(E)}{2}}.
 \label{Smtrpole1}
 \end{align}
 The  residue in the bound-state pole ${\cal A}_{s} \sim  C_{s}^{2}$  \cite{muk2023}, where $C_{s}$  is the ANC of the subthreshold  bound state. However, if we consider  the subthreshold bound state as the SR, ${\cal A}_{s} \sim \Gamma^{SR},\,$ where  $ \Gamma^{SR} \sim C_{s}^{2}$  is the width of the SR   and  $\Gamma(E) =  \Gamma^{SR}(E)+ \Gamma_{\gamma}(E)$  is the total resonance width of the SR.  
 A tiny radiative width of the SR decaying to the ground state, denoted by $\Gamma_{\gamma}(E)$, will be disregarded in the following discussion. 
Note that for energy values less than zero, $\Gamma^{SR}$ becomes zero due to the presence in it of the penetrability factor, causing the elastic scattering $S$-matrix element to exhibit a standard pole behavior,
 \begin{align}
 {\mathbb S} \stackrel{E \to -\varepsilon_{s}}{\sim}  \frac{{\cal A}_{s}}{E+ \varepsilon_{s}}.
 \label{Smtrpole2}
 \end{align}
 One can see that for small binding energy  $\varepsilon_{s}$  the pole  term  may significantly modify the behavior of the low-energy scattering wave function generating a new mechanism of the radiative capture, namely,  the radiative capture 
 ``continuum $\to$  the ground state"  occuring at positive energies of interacting nuclei but contributed  by the pole term of the $S$-matrix located at the negative energy.  This mechanism  can be called  the radiative capture to the ground state  through the SR.  Its amplitude  can be schematically written as
\begin{align}
M_{SR}  \sim  \frac{<\varphi_{f}\big| {\cal  H}_{el}^{L} \big| \sqrt{\Gamma_{SR}(E)}\,O>\, \sqrt{\Gamma_{SR}(E)}} {E+ \varepsilon_{s}   +  i\, \frac{\Gamma_{SR}(E)}{2}}
\label{MSRschem1}
\end{align} 
and  is described by the  pole diagram shown in Fig. \ref{fig_pole}. The diagram contains the form factor corresponding to the formation of the SR, resonance propagator and the form factor describing the radiative decay  of the SR to the ground state. It has a propagator similar to the real resonance propagator with the resonance energy $E_{R} >0$ replaced by
$-\varepsilon_{s}$.

It is now clear that in the presence of the subthreshold bound state the  amplitude $M_{NR}$
can be written as  the sum of two amplitudes:
\begin{align}
M_{NR}=  M'_{NR}  +  M_{SR},
\label{Mtot1}
\end{align}
where the nonresonant radiative capture amplitude $\,M'_{NR}\,$ is given by
\begin{align}
M'_{NR} \sim   <\varphi_{f}\big| {\cal  H}_{el}^{L}   \big| I - {\mathbb S}'\,O>.
\label{Mpr1}
\end{align}
Here  ${\mathbb S}'$  
is the elastic $S$-matrix element, which does not contain  the bound-state pole.
While the initial scattering wave function in the  nonresonant radiative transition amplitude $M'_{NR} $ contains both incident and outgoing waves, the SR  amplitude  $M_{SR}$  contains only  the outgoing scattered wave.
A distinctive trait  of the $R$-matrix approach  is the partitioning   of the radiative capture amplitude  into nonresonant  and SR  segments. The above presented schematic introduction can help the reader to better understand the phenomena of the SRs
playing a dominant role in the low-energy radiative capture  ${}^{12}{\rm C}(\alpha, \,\gamma){}^{16}{\rm O}$.

\subsection{Subthreshold resonances in the $R$-matrix formalism}
\label{SRRmatrix}

The main focus of this paper is to analyze the impact of the subthreshold and ground state ANCs on the low-energy $S$-factor for the radiative capture ${}^{12}{\rm C}(\alpha, \,\gamma){}^{16}{\rm O}.\,$  The ANC determines the normalization of the peripheral  amplitudes.  
This is the reason we focus on the radiative captures happening  in the external (channel) region ($r >R_{ch}$, where $R_{ch}$ is the channel radius).
This is why utilizing the $R$-matrix approach, which separates the configuration space into the internal and channel regions, is advantageous.  In the $R$-matrix formalism, the calculation of nonresonant (direct) capture to the ground state of the final nucleus only considers the external region, as the nonresonant capture in the internal region is included in the resonance capture amplitude.

There was no mention of the internal and external regions in Eqs. (\ref{scemMradcpt1})-(\ref{Mpr1}). Nonetheless, a rigorous approach dictates that the internal and channel regions be considered separately.
In the $R$-matrix aproach the total amplitude of the radiative transition to the ground state through the SR  is written as the sum of two parts,  internal  and channel:
\begin{align}
M_{SR}(E) = M_{SR}^{(int)}(E)  +  M_{SR}^{(ch)}(E),
\label{SRtot1}
\end{align}
where 
\begin{align}
M_{SR}^{(int)}(E) =& i\,e^{i(\sigma_{l_{s}}^{C} -  \delta_{l_{s}}^{hs}  )}\,\sqrt{2}\,(k_{\gamma}\,R_{ch})^{L+1/2}                           \nonumber\\
&\times \frac{\sqrt{\Gamma_{J_{s}}^{SR}(E)}\, \gamma_{\gamma\, J_{0}}^{J_{s}(int)}  }{E + \varepsilon_{s} +i\,\Gamma_{J_{s}}^{SR}(E)/2}
\label{MSRint1}
\end{align}
is the internal SR radiative transition amplitude and 
\begin{align}
M_{SR}^{(ch)}(E) =& i\,e^{i(\sigma_{l_{s}}^{C} -  \delta_{l_{s}}^{hs}  )}\,\sqrt{2}\,(k_{\gamma}\,R_{ch})^{L+1/2}                       \nonumber\\
&\times \frac{\sqrt{\Gamma_{J_{s}}^{SR}(E)}\, \gamma_{\gamma\, J_{0}}^{J_{s}\,(ch)}(E)}{E + \varepsilon_{s} +i\,\Gamma_{J_{s}}^{SR}(E) /2}
\label{MSRch1}
\end{align}
is the channel one.   Then the total amplitude  $M_{SR}(E)$  is given by
\begin{align}
M_{SR}(E) =& i\,e^{i(\sigma_{l_{s}}^{C} -  \delta_{l_{s}}^{hs}  )}\,\sqrt{2}\,(k_{\gamma}\,R_{ch})^{L+1/2}                          \nonumber\\
& \times \frac{\sqrt{\Gamma_{J_{s}}^{SR}(E)}\, \gamma_{\gamma\, J_{0}}^{J_{s}}(E)  }{E + \varepsilon_{s} +i\,\Gamma_{J_{s}}^{SR}(E)/2}.
\label{MSRtot1}
\end{align}
We explicitly indicated the total angular momentum of the SR  $J_{s}$  and the total angular momentum of the ground state $J_{0}=0$. Since in the case under consideration the channel spin is zero, $J_{s}=l_{s}$, where $l_{s}$ is the orbital angular momentum of the SR, it is enough to indicate only $J_{s}$ or $l_{s}$.  
Phase shift $\delta_{l_{s}}^{hs}$, determined by
\begin{align}
e^{-2\,i\,\delta_{l_{s}}^{hs} }= \frac{G_{l_{s}}(k,\,R_{ch}) - i\,F_{l_{s}}(k,\,R_{ch}) }{G_{l_{s}}(k,\,R_{ch}) + i\,F_{l_{s}}(k,\,R_{ch})} ,
\label{deltahs1}
\end{align}
and $\,\sigma_{l_{s}}^{C}$  are the hard-sphere and Coulomb  scattering phase shifts in the partial wave $l_{s}$, respectively.  Functions $\,F_{l_{s}}(k,\,r)$ and $G_{l_{s}}(k\,r)$  are the regular and irregular Coulomb solutions,
$\Gamma_{J_{s}}^{SR}(E)$   and  $\gamma_{\gamma\,J_{0}}^{J_{s}}(E)$  are  the observable  width  and  the reduced  radiative width amplitude of  the SR, respectively.

Even though the SR is situated at negative energy as a subthreshold bound state, its resonance width is specified at $E>0$ and vanishes at
$E \leq 0$:
\begin{align}
&\Gamma_{J_{s}}^{SR}(E) =   P_{l_{s}}(k, R_{ch})\,[\gamma_{l_{s}}^{SR}(R_{ch})]^{2},
\label{GammaSR1}
\end{align}
where at $E>0$  the penetrability factor is
\begin{align}
P_{l_{s}}(k,\,R_{ch})&= \frac{k\,R_{ch}}{F_{l_{s}}^{2}(k,\,R_{ch}) +   G_{l_{s}}^{2}(k,\,R_{ch}) } \\
&=  \frac{k\,R_{ch}}{\big|O_{l_{s}}(k,\,R_{ch})\big|^{2}},
\label{Pntrfctr1}
\end{align}
 $O_{l_{s}}$ is the outgoing Coulomb scattered wave\footnote{See Ref. \cite{mukblokh2022}  for details on how factor $e^{\pi\,\eta\,{\rm sign}({\rm Re}k)/2}$  appears.}:
\begin{align}
 O_{l_{s}}(k,\,r)&=  e^{-i\,\sigma_{l_{s}}^{C}}\,[G_{l_{s}}(k,\,R_{ch})+ i\,F_{l_{s}}(k,\,R_{ch})] 
\nonumber\\
&=e^{-i\,\pi\,l_{s}/2}\,e^{\pi\,\eta\,{\rm sign}({\rm Re}k)/2}\,W_{-i\,\eta,\, l_{s}+1/2}(-i\,2\,k\,r), 
\label{OlFlGl1}
\end{align} 
\begin{align}
{\rm sign}(x)= \Biggl \{ \begin{array}{rcl}
1 & \mbox{for}
& x>0, \\ 0 & \mbox{for} & x=0, \\
-1 & \mbox{for} & x<0,
\label{sign1}
\end{array}
\end{align}
$W_{-i\,\eta,\, l_{s}+1/2}(-i\,2\,k\,r)$ and $\eta$ are the Whittaker function and the Coulomb parameter for $k>0$.
In addition, 
\begin{align}
& [\gamma_{l_{s}}^{SR}(R_{ch})]^{2}= \frac{(\hbar\,c)^{2}}{\mu}\,\frac{W_{-\eta_{s},\,l_{s}+1/2}^{2}(2\,\kappa_{s}\,R_{ch})}{R_{ch}}\,C_{l_{s}}^{2} 
\label{gSR1}
\end{align}
is the 
observable
reduced width of the SR, see Appendix \ref{Appendix1} and Refs.\cite{muk2023,muktribble1999}.
 $W_{-\eta_{s},\,l_{s}+1/2}(2\,\kappa_{s}\,R_{ch})$  is the Whittaker function describing the radial behavior  of the bound-state wave function in  the external region,  $\kappa_{s}= \sqrt{2\,\mu\,\varepsilon_{s}}$ is the wave number of the subthreshold bound  state,  $\,\mu$ is the reduced mass of the interacting nuclei expressed  in MeV,  $\eta_{s}$  is  the $\alpha-{}^{12}{\rm C}$ Coulomb parameter of  the subthreshold bound state,   $C_{l_{s}}$  is the ANC  of the subthreshold bound state expressed in fm${}^{-1/2}$. 
 We explicitly inserted $(\hbar\,c)^2$ in Eq. (\ref{gSR1}) to validate that $ [\gamma_{l_{s}}^{SR}(R_{ch})]^{2}$ is measured in MeV.
   Equation (\ref{gSR1}) clearly shows that  the reduced width of the SR  depends only on the  ANC $C_{l_{s}}$  and the channel radius radius $R_{ch}$.   Hence the overall normalization  of  $\sqrt{\Gamma_{J_{s}}^{SR}(E)}$  is expressed in terms of the ANC of the subthreshold bound state $C_{l_{s}}.\,$  
 
 Equation (\ref{MSRtot1})   illustrates
 that  the SR is a resonance located  at the negative energy of  $-\varepsilon_{s}.\,$   $\, \gamma_{\gamma\, J_{0}}^{J_{s}\,(int)}(E)$ and $ \gamma_{\gamma\, J_{0}}^{J_{s}\,(ch)}(E)$  
 in Eqs.  (\ref{MSRint1})   and  (\ref{MSRch1})  are the  internal and channel radiative width amplitudes of the SR  with the total one
\begin{align}
\gamma_{\gamma\, J_{0}}^{J_{s}\,}(E) = \gamma_{\gamma\, J_{0}}^{J_{s}\,(int)}  + \gamma_{\gamma\, J_{0}}^{J_{s}\,(ch)}(E).
\label{ggtot1}
\end{align} 
In the $R$-matrix approach  $\,\gamma_{\gamma\, J_{0}}^{J_{s}\,(int)} $   is  a real constant,   while  $\gamma_{\gamma\, J_{0}}^{J_{s}\,(ch)}(E)$  is complex and energy-dependent.  

An explicit expression for   $\gamma_{\gamma\, J_{0}}^{J_{s}\,(ch)}(E)$   is 
\begin{align}
& \gamma_{\gamma\, J_{0}}^{J_{s}\,(ch)}(E) = C_{l_{0}}\, {\cal D}_{L}(E)\,\sqrt{P_{l_{s}}(k, R_{ch})}\,\gamma_{l_{s}}^{SR}(R_{ch})\,
 \nonumber\\
&\times\,<l_{s}\,0\,\,L\,0 \big|l_{0}\,0>\,U(L\,l_{0}\, s_{s}\,J_{s}; l_{s}\,J_{0})\,{\cal J}_{ch}(E),
\label{gammagSRch1}         
\end{align}
where
\begin{align}
 {\cal D}_{ L}(E)=  &  \sqrt{\frac{1}{2\,E}}\, \frac{1}{ (2\,L+1)!!}\, \sqrt{\frac{(L+1)\,(2\,L+1)}{L}}\,
    \nonumber\\
& \times  Z_{eff}( L)\,\sqrt{k\,R_{ch}}
\label{DdL1}
\end{align}
and  (in a  concise form)
 \begin{align}
 {\cal J}_{ch}(E)= & \frac{1}{R_{ch}^{L+1}}\,\int_{R_{ch}}^{\infty} {\rm d}r\,r^{L}\,W_{-\eta_{0},\,l_{0}+1/2}(2\,\kappa_{0}\,r)  \nonumber\\
&\times e^{- i\, \delta_{l_{s}}^{hs}  } O_{l_{s}}(k,r).
 \label{calJch1}
  \end{align}
  $W_{-\eta_{0},\,l_{0}+1/2}(2\,\kappa_{0}\,R_{ch})$
 is the Whittaker function describing the external radial behavior of the   $\alpha-{}^{12}{\rm C}$  bound-state wave function in the ground state.
Subscript $s$ denotes the continuum state corresponding to the SR, while subscript $0$ refers to the ground state of ${}^{16}{\rm O}$.
$l_{0}=0,\,$ $\,\kappa_{0}=\sqrt{2\,\mu\,\varepsilon_{0}}\,$  and $\varepsilon_{0}=7.162$ MeV  are the $\alpha-{}^{12}{\rm C}$ orbital angular momentum, bound-state wave number and the binding energy of the  ${}^{1 6}{\rm O}$ ground state,  respectively.  $\eta_{0}$  is  the $\alpha-{}^{12}{\rm C}$ Coulomb parameter of  the ground bound state.
The quantity 
\begin{align} 
Z_{eff}(L)=e\, \mu^{L}\,\Big(\frac{Z_{\alpha}}{m_{\alpha}^{L}}  + (-1)^{L}\,\frac{Z_{{}^{12}{\rm C}}}{ m_{{}^{12}{\rm C}}^{L}}\Big)
\label{ZeffL}
\end{align}
is the effective charge of the system for the transition with multipolarity $L,\,$  
and $e$ is the proton charge.

Rewriting  this matrix element in terms of the regular $F_{l_{s}}(k,r)$ and irregular $G_{l_{s}}(k,r)$  Coulomb  solutions  we get \cite{BarkerKajino}
\begin{align}
{\cal J}_{ch}(E) =& W_{-\eta_{0},\,l_{0}+1/2}(2\,\kappa_{0}\,R_{ch})\, \sqrt{F_{l_{s}}^{2}(k,\,R_{ch}) +  G_{l_{s}}^{2}(k,\,R_{ch}) }
\nonumber\\
&\times \,J_{L\,l_{s}\,l_{0}}(E),
\label{JchJL1}
\end{align}
where  
\begin{align}
J_{L\,l_{s}\,l_{0} }(E)= & J_{ 2\, L\,l_{s}\,l_{0} }(E)          \nonumber\\
& +  i\,\frac{  F_{  l_{s}}(k,\,R_{ch})\,G_{l_{s}}(k,\,R_{ch}) }{ F_{l_{s}}^{2}(k,\,R_{ch}) +  G_{l_{s}}^{2}(k,\,R_{ch}) }\,J_{1\,L\,l_{s}\,l_{0}}(E),
\label{JLlsl01}
\end{align}
\begin{align}
J_{1\,L\,l_{s}\,l_{0}}(E)= & \frac{1}{R_{ch}^{L+1}}\,\int\limits_{R_{ch}}^{\infty}\,{\rm d}r\,r^{L}\,\frac{W_{-\eta_{0},\,l_{0}+1/2}(2\,\kappa_{0}\,r)}{W_{-\eta_{0},\,l_{0}+1/2}(2\,\kappa_{0}\,R_{ch})}\,
\nonumber\\
&\times \Big[\frac{F_{l_{s}}(k,\,r)}{F_{l_{s}}(k,\,R_{ch})}  -  \frac{G_{l_{s}}(k,\,r)}{G_{l_{s}}(k,\,R_{ch})} \Big],
\label{JLlslopr1}
\end{align}
and
\begin{align}
&J_{2\,L\,l_{s}\,l_{0}}(E)=  \frac{1}{R_{ch}^{L+1}}\,\int\limits_{R_{ch}}^{\infty}\,{\rm d}r\,r^{L}\,\frac{W_{-\eta_{0},\,l_{0}+1/2}(2\,\kappa_{0}\,r)}{W_{-\eta_{0},\,l_{0}+1/2}(2\,\kappa_{0}\,R_{ch})}\,
\nonumber\\
&\times \Big[\frac{ F_{l_{s}}(k,\,r)\,F_{l_{s}}(k,\,R_{ch}) + G_{l_{s}}(k,\,r)\,G_{l_{s}}(k,\,R_{ch})}{F_{l_{s}}^{2}(k,\,R_{ch}) +  G_{l_{s}}^{2}(k,\,R_{ch}) }   \Big].
\label{JLlslopr1}
\end{align}

Then
\begin{align}
\gamma_{\gamma\, J_{0}}^{J_{s}\,(ch)}(E)= & C_{l_{0}}\,\sqrt{k\,R_{ch}}\, {\cal D}_{L}(E)\,\gamma_{l_{s}}^{SR}(R_{ch})\,
 \nonumber\\
&\times\,<l_{s}\,0\,\,L\,0 |l_{0}\,0>\,
U(L\,l_{0}\, s_{s}\,J_{s}; l_{s}\,J_{0})            \nonumber\\
& \times W_{-\eta_{0},\,l_{0}+1/2}(2\,\kappa_{0}\,R_{ch})\,J_{L\,l_{s}\,l_{0} }(E).
\label{gammagSRch1}         
\end{align}
$C_{l_{0}}$  is  the ANC of the ${}^{1 6}{\rm O}$ ground state, $Z_{i}$ is the  number of the protons  in nucleus $i$,  $\,k_{\gamma} =E+ \varepsilon_{0}$ is the momentum of the emitted photon,  $k=\sqrt{2\,\mu\,E},\,$   $<l_{s}\,0\,\,L\,0 |l_{0}\,0>$  is the  Clebsch-Gordan coefficient, 
$U(L\,l_{0}\, s_{s}\,J_{s}; l_{s}\,J_{0})\,$
is the  Racah  coefficient,  $s_{s} =0$ is the $\,\alpha-{}^{12}{\rm C}\,$  channel spin in the subthreshold state.   

For the bound states $J_{L\,l_{s}\,l_{0} }$ is given by  \cite{BarkerKajino}
\begin{align} 
J_{L\,l_{s}\,l_{0} }= & \frac{1}{R_{ch}^{L+1}}\,\int\limits_{R_{ch}}^{\infty}\,{\rm d}r\,r^{L}\,\frac{W_{-\eta_{0},\,l_{0}+1/2}(2\,\kappa_{0}\,r)}{W_{-\eta_{0},\,l_{0}+1/2}(2\,\kappa_{0}\,R_{ch})}        \nonumber\\
&\times \frac{W_{-\eta_{s},\,l_{s}+1/2}(2\,\kappa_{s}\,r)}{W_{-\eta_{s},\,l_{s}+1/2}(2\,\kappa_{s}\,R_{ch})}.
\label{JLlsbs1}
\end{align}
The channel radiative width pertaining to the bound state is given by
\begin{align}
& \gamma_{\gamma\, J_{0}}^{J_{s}\,(ch)}(-\varepsilon_{s}) = \sqrt{\mu}\,
 Z_{eff}(L)\,R_{ch}\,\frac{1}{ (2\,L+1)!!}                   \nonumber\\
&\times \sqrt{\frac{(L+1)\,(2\,L+1)}{L}}\,C_{l_{0}}\, W_{-\eta_{0},\,l_{0}+1/2}(2\,\kappa_{0}\,R_{ch})    \nonumber\\
&\times <l_{s}\,0\,\,L\,0 |l_{0}\,0>\,
U(L\,l_{0}\, s_{s}\,J_{s}; l_{s}\,J_{0}) \,
\gamma_{J_{s}}(R_{ch})
J_{L\,l_{s}\,l_{0} },  
\label{ggchbs1}
\end{align}
where $\gamma_{J_{s}}(R_{ch})$  is the reduced width of the subthreshold  bound state, which is expressed in terms of the subthreshold ANC $C_{J_{s}}$.

We see that  the   normalization of the channel radiative width is  expressed in terms of the subthreshold  ANC $C_{l_{s}}$  and the ANC  $C_{l_{0}}$  of the ground state of ${}^{16}{\rm O}.\,$   It explains why these  ANCs  govern the behavior of the  low-energy amplitude  $M_{SR}^{(ch)}(E)$   describing the channel resonance radiative transition to the ground state  through the SR.   

The radiative width of the SR is expressed in  terms of   $\gamma_{\gamma\, J_{0}}^{J_{s}\,}(E) $  as
\begin{align}
\Gamma_{\gamma\, J_{0}}^{J_{s}\,}(E) = &2\,(k_{\gamma}\,R_{ch})^{2\,L+1}\, \big|\gamma_{\gamma\, J_{0}}^{J_{s}\,}(E)  \big|^{2}   
\label{Gggg1}   \\
= &2\,(k_{\gamma}\,R_{ch})^{2\,L+1}\, \big|\gamma_{\gamma\, J_{0}}^{J_{s}\,(int)}  + \gamma_{\gamma\, J_{0}}^{J_{s}\,(ch)}(E) \big|^{2} .
\label{Gggg2}
\end{align}

If  $\big| \gamma_{\gamma\, J_{0}}^{J_{s}\,(int)} \big |   >>  \big| \gamma_{\gamma\, J_{0}}^{J_{s}\,(ch)}(E)  \big| $, then  the energy  dependence  of  $\Gamma_{\gamma\, J_{0}}^{J_{s}\,}(E)$   is determined
by the dimensionless factor $(k_{\gamma}\,R_{ch})^{2\,L+1}$  and  we can write
\begin{align}
\Gamma_{\gamma\, J_{0}}^{J_{s}\,}(E)= \Big( \frac{E+ \varepsilon_{0}}{\varepsilon_{0} -  \varepsilon_{s}}   \Big)^{2\,L+1}\,\Gamma_{\gamma\, J_{0}}^{J_{s}\,}(-\varepsilon_{s}).
\label{GgE1}
\end{align}
 This equation allows one to find the 
radiative width  $\Gamma_{\gamma\, J_{0}}^{J_{s}\,}(E)$  at  positive energies if  (it is often the case)  $\Gamma_{\gamma\, J_{0}}^{J_{s}\,}(-\varepsilon_{s})$ is known experimentally.

 We remind that $\,\gamma_{\gamma\, J_{0}}^{J_{s}\,(ch)}(E)$  can be explicitly calculated.  If its energy dependence  is  weak, then Eq. (\ref{GgE1})  can be  a good approximation even for the nonnegligible  channel part.  Then, in view of Eq. (\ref{Gggg2}),   
 one can find  $\gamma_{\gamma\, J_{0}}^{J_{s}\,(int)}$:
 \begin{align}
\gamma_{\gamma\, J_{0}}^{J_{s}\,(int)}  =& - {\rm Re}  \gamma_{\gamma\, J_{0}}^{J_{s}\,(ch)}(E)  \nonumber\\
& \pm \sqrt{  \frac{\Gamma_{\gamma\, J_{0}}^{J_{s}\,}(E)}{2\,(k_{\gamma}\,R_{ch})^{2\,L+1}} - \big[{\rm Im}  \gamma_{\gamma\, J_{0}}^{J_{s}\,(ch)}(E)\big]^{2}  }.
\label{ggint2}
\end{align}

\subsection{Direct radiative capture}
\label{Nonrescapt1}

There is another radiative process where the subthreshold bound states contribute: the cascade radiative capture to the ground state  through the subthreshold bound  states. It is a two-step process, which  begins with the first direct radiative capture to the subthreshold bound state, which is then followed by its decay to the ground state through photon emission.
 The direct capture amplitude  to the subthreshold state, which is the first part of the  cascade transition to the ground state,  is given by
\begin{align}
 {\cal M}_{DC(s)}(E)  = & e^{i(\sigma_{l_{i}}^{C} - \delta_{l_{i}}^{hs}) }\, {\cal D}_{L}(E)\, (k_{\gamma}\,R_{ch})^{L+1/2}\,C_{l_{s}}\,   \nonumber\\
&\times <l_{i}\,0\,\,L\,0 |l_{s}\,0>                                                                                                               
\,U(L\,l_{s}\, s_{i}\,J_{i}; l_{i}\,J_{s}) \,{\cal J}_{is}(E),   
\label{MDCSR1}
\end{align}
with
 \begin{align}
 {\cal J}_{is}(E)  = & \frac{1}{R_{ch}^{L+1}}\,\int_{R_{ch}}^{\infty} {\rm d}r\,r^{L}\,W_{-\eta_{s},\,l_{s}+1/2}(2\,\kappa_{s}\,r)   \nonumber\\
&\times \big[ e^{-i\,\delta_{l_{i}}^{hs} }\,I_{l_{i}}(k,r)  - e^{i\,\delta_{l_{i}}^{hs} }O_{l_{i}}(k,r)  \big].
 \label{Jis1}
 \end{align}
Here  $\,l_{i}$  is the $\alpha-{}^{12}{\rm C}$  orbital angular momentum  in the continuum, $s_{i}=0$ is their channel spin. 

Rewriting  ${\cal J}_{is}(E) $   in terms of the  regular and  irregular Coulomb solutions  we get
\cite{BarkerKajino}
\begin{align}
{\cal J}_{is}(E)= &  2\,i\,W_{-\eta_{s},\,l_{s}+1/2}(2\,\kappa_{s}\,R_{ch})                     \nonumber\\
& \times \frac{F_{l_{i}}(k,\,R_{ch})\,G_{l_{i}}(k,\,R_{ch}) }{\sqrt{F_{l_{i}}^{2}(k,\,R_{ch})  +   G_{l_{i}}^{2}(k,\,R_{ch})}}\,    
J_{1\,L\,l_{i}\,l_{s}}(E).
\label{JisJpr1}
\end{align}
 
Equation (\ref{MDCSR1}) clarifies why in the $R$-matrix approach the  overall normalization of the direct capture amplitiude  is determined by the  ANC of the final bound state formed as the result of the direct radiative capture \cite{mukblokh2022}. 

We  explicitly showed only the amplitude of the first step of the cascade  radiative capture. 
For the subthreshold  case under consideration, the cascade  transitions through   $(7.12\, {\rm MeV}, 1^{-})\,$  and  $(6.92\, {\rm MeV}, 2^{+})\,$ subthreshold bound states are significantly weaker than the resonant captures to the ground state when these bound  states reveal themselves as the SRs (see Fig. \ref{tot_Sfctrs}  below). 
In addition to direct capture to the subthreshold bound state, it is necessary to consider
the direct   ``continuum to ground  state" capture.  The amplitude of this transition is similar  to  the amplitude  $\, {\cal M}_{DC(s)}(E)\, $ (see Eq. (\ref{MDCSR1}))  and is given by
\begin{align}
 {\cal M}_{DC(0)}(E)  = & e^{i(\sigma_{l_{i}}^{C} - \delta_{l_{i}}^{hs}) }\, {\cal D}_{L}(E)\,(k_{\gamma}\,R_{ch})^{L+1/2}\, C_{0}
 \nonumber \\ 
 & \times    
<l_{i}\,0\,\,L\,0 |l_{0}\,0>\,U(L\,l_{0}\, s_{i}\,J_{i}; l_{i}\,J_{0})\, 
{\cal J}_{i0}(E).
\label{MDGR1}
\end{align}  
The formula for ${\cal J}_{i0}(E)$ can be derived from ${\cal J}_{is}(E)$ (refer to Eq. (\ref{Jis1})) by substituting subscript $s$ with $0$.
Note that  the only unknown quantity  in Eq. (\ref{MDGR1}) is  the ANC  $\,C_{0}\,$ of the ground  state of 
${}^{16}{\rm O}$.   This ANC can be positive or negative \footnote{By definition, the ANC is a real quantity.} determining the sign of  $ {\cal M}_{DC(0)}(E).\, $   For $l_{i}=l_{s}$   the direct capture amplitude  $ {\cal M}_{DC(0)}(E)$  interferes with  the  amplitude $M_{SR}(E)$  for  the resonance capture to the ground state through the SR. 
The sign of  Eq. (\ref{gammagSRch1}) and, hence, of  $M_{SR}(E)$  is controlled  by the product  $C_{l_{0}}\,C_{l_{s}}$. If we fix the sign of the subthreshold ANC $C_{l_{s}}$,  then  the nonresonant amplitude $ {\cal M}_{DC(0)}(E) $  and the channel resonance radiative   amplitude $M_{SR}^{(ch)}(E)$  are normalized in terms of the  same ANC, $C_{0}$.   Such a normalization is physically transparent:  both amplitudes describe peripheral processes and, hence, contain the tail  of the nuclear overlap function of the ground state, whose normalization is given by the corresponding ANC.

 For the interfering direct capture to the ground state and the resonance  capture to the ground state through the SR, 
 it is convenient  to write  the  sum of the interfering amplitudes as
 \begin{align}
M_{SR}(E) +  {\cal M}_{DC(0)}(E)= &  [ M_{SR}^{(ch)}(E)  +   {\cal M}_{DC(0)}(E)]         \nonumber\\
& +  M_{SR}^{(int)}(E).
\label{MDRSRSint1}
\end{align}
 The relative sign of the sum in the brackets is  well determined, while  the sign  of  $M_{SR}^{(int)}(E)$, which contains  $\gamma_{\gamma\, J_{0}}^{J_{s}\,(int)},\, $   is the fitting parameter. 
 
The equations in this section shed light on the reason behind the extensive effort dedicated to determining the ANCs of the $\alpha$-particle removal from the subthreshold bound states of ${}^{16}{\rm O}$  (see \cite{deBoer}). The different experimental and theoretical methods of determining the ANCs are discussed in \cite{mukblokh2022}.

\subsection{Resonant transition}

Since the main purpose of this paper is to investigate  the role of the  subthreshold  ANCs, we only briefly discuss  the resonance capture to the ground state of ${}^{16}{\rm O}$.  Equations  for the resonance capture can be obtained  from the corresponding  equations for  resonance capture through the SR  by replacing  $s \to i,\,$  $-\varepsilon_{s}  \to  E_{R}$, where   $E_{R}$  is the resonance energy\footnote{The $R$-matrix approach deals  only with  real resonance energies.},  $\Gamma_{J_{s}}^{SR}(E)   \to  \Gamma_{J_{i}}(E)$  and   $ \gamma_{\gamma\, J_{0}}^{J_{s}}(E)   \to \gamma_{\gamma\, J_{0}}^{J_{i}}(E).\,$   Here   $\Gamma_{J_{i}}(E)$   is  the resonance width,  $ \gamma_{\gamma\, J_{0}}^{J_{i}}(E) $   is the  resonance radiative width amplitude,  
which  can be split
into the internal and channel parts;  $l_{i}$  is $\alpha-{}^{12}{\rm C}$   resonance  orbital angular momentum.  For   $l_{i}= l_{s}$   the   resonance  amplitude interferes with the SR  and direct capture amplitudes.
Adding  higher  resonances  interfering with the SR requires  employing the multilevel $R$-matrix equations \cite{deBoer,azuma}.

\subsection{\bf{ANC of the ground state of $\mathbf{{}^{16}{\rm O}}$}}

Here we discuss the impact of the ground state ANC $C_{l_{0}}$.   This ANC  enters  the channel radiative widths of the SR and the resonance,  cascade  transition  and   direct  capture amplitudes to the ground  state.  Owing to the large binding energy of the ground state of the system $\,\alpha + {}^{12}{\rm C}\,$   ($\varepsilon_{0}= 7.16$ MeV),  the internal  part of the radiative width  of the SR   $ \gamma_{\gamma\, J_{0}}^{J_{s}\,(int)}(E)$ and  the resonance internal radiative width $\,\gamma_{\gamma\, J_{0}}^{J_{i}\,(int)}(E)$ 
 are dominant compared  to the corresponding channel ones containing $C_{l_{0}}$. Besides, for the $E1$ transition the channel part is suppressed  because the effective charge  $Z_{eff}(1)$  is very small  ( $Z_{eff}(1)/e= -0.00097$).  It significantly diminishes the role of the channel radiative widths. For the same reason, the $E1$ direct capture to  the ground state is suppressed.  The numerical evidence provided below indicates that the ground-state ANC has no influence on the $E1$ $S(300\,{\rm keV})$-factor at low energies.
 
 Also  the  cascade transitions  to the ground state  are small compared to the SR and resonance transitions.  
It makes the impact of the ground-state ANC less important than that of the ANCs of the subthreshold states $1^{-}$ and $2^{+}$, but not negligible  due to the interference  of the  $2^{+}$  SR, resonance and the direct transition to the ground state.  
The role of $C_{l_{0}}$   on the low-energy  $S$-factor,  especially, $S(300\,{\rm keV})$, is analyzed through numerical calculations below.
Two sets of subthreshold ANCs from the low end \cite{deBoer} and high end \cite{Blokh2023} are used in this paper for systematic comparison.
The latter set was determined  using the extrapolation procedure of the elastic scattering phase shifts. However,  the ground-state ANC was not determined using the extrapolation method because it  is located quite far from the threshold  (the binding  energy of the $\alpha$-particle in the ${}^{16}{\rm O}$  ground state is $7.16$ MeV). The  values of the ground-state ANC published in the literature vary significantly~\cite{Shen}.  
 In this paper,  we explore two values of this  ANC: $58$ fm${}^{-1/2}$  \cite{deBoer} and  $337$  fm${}^{-1/2}$ \cite{Shen}.
 The  ground-state ANC of  $337 \pm 45$  fm${}^{-1/2}$  found  in \cite{Shen} from the heavy-ion induced transfer reaction requires a higher value of $(1.55 \pm 0.09) \times 10^{5}$  fm${}^{-1/2}$  for the  ANC of the $(2^{+},\, 6.92$ MeV) excited state to reconcile with the $S$-factor from \cite{deBoer}  calculated for the ground-state ANC of $58$  fm${}^{-1/2}$.  This value of the $2^{+}$ subthreshold ANC is close but slightly higher than the  ANC for the $6.92$ MeV state from \cite{Blokh2023} (see Table \ref{Table_ANCs}).  

\subsection{Subthreshold ANCs}
\label{SubANCs1}

One of the main goals of this paper  is to check the impact of the ANCs of the $\alpha$-particle removal  from  the subthreshold states $1^{-}$ and $2^{+}$ of ${}^{16}{\rm O}^{*}$.
The ANCs  of these subthreshold bound states  available in the literature 
have been recently tabulated in Ref. \cite{deBoer}.  This revealed 
large discrepancies between  the  ANCs determined by different techniques, suggesting further efforts are needed  to pinpoint the ANCs of the two near-threshold bound states  in  ${}^{16}{\rm O}$.  
 The  ANC  values  for the channels ${}^{16}{\rm O}^{*}  \to  \alpha + {}^{12}{\rm C}$ obtained by various methods 
are also compared in Table \ref{Table_ANCs}.  
\begin{table*}[htb]
\caption{ANC $C_l$ values in fm$^{-1/2}$ for $^{16}$O$^*(J^\pi)\to \alpha+^{12}$C(g.s.)}
\begin{center}
\begin{tabular}{|c|c|c|c|c|}
\hline
$C_0$; $J^{\pi}=0^+$  & $C_3$; $J^{\pi}=3^-$ & $C_2$; $J^{\pi}=2^+$  & $C_1$; $J^{\pi}=1^-$ & References \\
$\varepsilon=1.113$ MeV & $\varepsilon=1.032$ MeV & $\varepsilon=0.245$ MeV & $\varepsilon=0.045$ MeV & \\
\hline
- & - & (1.11$\pm$0.10)$\times 10^5$ & (2.08$\pm$0.19)$\times 10^{14}$ & \cite{Brune} \\
-                              &  -                             & (1.40$\pm$0.42)$\times 10^{5}$ & (1.87$\pm$0.32)$\times 10^{14}$ & \cite{Balhout} \\
-                              & -                              & (1.44$\pm$0.26)$\times 10^{5}$ & (2.00$\pm$0.69)$\times 10^{14}$ & \cite{Oulevsir} \\
(1.56$\pm$0.09)$\times 10^{3}$ & (1.39$\pm$0.08)$\times 10^{2}$ & (1.22$\pm$0.06)$\times 10^{5}$ & (2.10$\pm$0.14)$\times 10^{14}$ & \cite{Avila} \\
-                              & -                              & 0.213$\times 10^{5}$ & 1.03$\times 10^{14}$ & \cite{Orlov1} \\
0.4057$\times 10^{3}$           & -                              & 0.505$\times 10^{5}$ & 2.073$\times 10^{14}$ & \cite{Orlov2} \\
- & -& (1.10$-$1.31)$\times 10^{5}$ & 2.21(0.07)$\times 10^{14}$ & \cite{Sparen} \\
(0.64$-$0.74)$\times 10^{3}$ & (1.2$-$1.5)$\times 10^{2}$ & (0.21$-$0.24)$\times 10^{5}$ & (1.6$-$1.9)$\times 10^{14}$ & \cite{Ando} \\
0.293$\times 10^{3}$ & - & - & - & \cite{Orlov3} \\
1.34$\times 10^{3}$ & 1.22$\times 10^{2}$ & (0.98$-$1.07)$\times 10^{5}$  & (1.83$-$1.84)$\times 10^{14}$   & \cite{Hebborn} \\
$1.56 \times 10^{3}$ & $1.39 \times 10^{2}$ & $1.14 \times 10^{5}$  & $2.08 \times 10^{14}$   & \cite{deBoer} \\
(0.886$-$1.139)$\times 10^{3}$ & - & - & - & \cite{BKMS5} \\
- & (2.17$\pm$0.05)$\times 10^{2}$ & (1.42$\pm$0.05)$\times 10^5$ & (2.27$\pm$0.02)$\times 10^{14}$ & \cite{Blokh2023} \\
\hline 
\end{tabular}
\end{center}
\label{Table_ANCs}
\end{table*}

It is not the goal of this paper to verify the reliability and uncertainty of  each  subthreshold ANC published in the literature. 
In order to assess the effects of the subthreshold ANCs, as previously mentioned, we will be analyzing two sets:
one represents the lower end of the published subthreshold ANCs and is taken from \cite{deBoer},  while the second  set representing the high-end of the subthreshold ANCs  was reported recently ( the last two rows in Table  \ref{Table_ANCs}, which are   taken from   \cite{BKMS5} and \cite{Blokh2023}). 

The ANCs obtained in \cite{BKMS5}  and  \cite{Blokh2023}  were obtained by the extrapolation of the experimental phase shifts  in the corresponding partial waves to the subthreshold bound-state poles located at negative energies.
Unfortunately, uncertainties of the experimental phase shifts are unknown. Therefore, in this paper, in order to estimate uncertainties  of  the ANCs obtained by the extrapolation method we  assumed  $5\%$ uncertainty in the experimental  phase shifts. 

To be consistent  we included in our calculations all four subthreshold states in ${}^{16}{\rm O}$,  although we understand that only $1^{-}$ and $2^{+}$ give the dominant contribution.
Note that the uncertainty of the ANC for the  $0_{2}^{+}$ state determined by applying the extrapolation method 
is significantly higher than for the three weaker bound subthreshold states presented in Table  \ref{Table_ANCs}.  The reason is that the farther  the threshold state from the pole that corresponds to the bound state in the energy plane,  the lower is the accuracy of the ANC obtained by the extrapolation of the elastic-scattering phase shifts to the corresponding bound-state pole.

\subsection{\bf{Radiative width amplitudes  of the subthreshold  resonances $1^{-}$  and $\,2^{+}$}}
\label{Radwidths1}

The  amplitudes $M_{SR}(E)$  of the radiative capture  ${}^{12}{\rm C}(\alpha, \,\gamma){}^{16}{\rm O}$   
through the 
$1^{-}$ and $2^{+}$
SRs  give the dominant contribution  to the low-energy $S$-factors.  These amplitudes 
depend on the  resonance widths of  SRs  $\Gamma_{J_{s}}^{SR}(E), $  which can be calculated explicitly,  and the radiative width amplitudes  $\,\gamma_{\gamma\, J_{0}}^{J_{s}\,}(E)\,$  of the SRs. 
Before presenting the results of the low-energy $S$-factor calculations, we will review the  SR radiative widths amplitudes.

\subsubsection{Subthreshold resonance  $1^{-}$  }
\label{SR1}
First, we consider the radiative width amplitude  for the SR $1^{-}$.  Owing to the fact  the ratio 
$Z_{i}/m_{i}$ for the $\alpha$-particle and  ${}^{12}{\rm C}$ is practically the same, the  effective charge for the $L=1$ transition is  very small  and the channel radiative width amplitude  $ \gamma_{\gamma\, J_{0}}^{J_{s}\,(ch)}(E)$  should be negligible compared to the internal  counterpart.\footnote[3]{Each of involved nuclei, ${}^{4}{\rm He}$ and ${}^{12}{\rm C}$  have  $Z_{i}=N_{i}$, where $N_{i}$ is the number of neutrons in nucleus $i$.  For systems with such nuclei  the isospin projection $T_{3}=0$ and the isospin selection rule  requires change of the isospin of the system by one. Since it is not fulfilled in our case, the considered $E1$ transition $1^{-}  \to 0^{+}$ is called isospin forbidden \cite{AisenbergGreiner,Baye}. However, the suppression is not very strict because of the mass difference between the protons and neutrons and the residual Coulomb effects.}  
In Table  \ref{RDwa1}  are shown  the low-energy dependences of the  channel and  internal radiative width amplitudes for the SR $1^{-}.$  Just as anticipated, the inner component is unchanging. 
   We can see that  the internal part is significantly larger than the channel part, leading to the validity of Eq. (\ref{GgE1}).  The experimental value  of the radiative width of the subthreshold bound state  $1^{-}$
decaying to the ground state  is  $\Gamma_{\gamma\, 0}^{1}(-0.045\,{\rm MeV})= (5.5 \pm 0.3)\times 10^{-8}$  MeV  \cite{TUNL}.  Hence the internal radiative width amplitude can be calculated using Eqs. (\ref{GgE1}) and (\ref{ggint2}).  A $6\%$  experimental uncertainty  of $\Gamma_{\gamma\, 0}^{1}(-0.045\,{\rm MeV})$ propagates into the $3\%$ uncertainty of $\gamma_{\gamma\, 0}^{1}\,(int)$.
$\gamma_{\gamma\, 0}^{1\,(p)}\,(int)$   ( $\gamma_{\gamma\, 0}^{1\,(m)}\,(int)$)  in Table \ref{RDwa1}   corresponds to the higher (lower) solution of Eq. (\ref{ggint2}).   For practical calculations we use 
$\gamma_{\gamma\, 0}^{1\,(p)}\,(int)=  
(1.46 \pm 0.05) \times 10^{-3}
$  MeV${}^{1/2}$.
It should be noted that even for the high subthreshold ANC (the last row of Table \ref{Table_ANCs})  and  the high ground-state ANC of $337$ fm${}^{-1/2}$ \cite{Shen}  the channel radiative width amplitude  $\gamma_{\gamma\, 0}^{1\,(ch)}(300\,{\rm keV})= 2.06 \times10^{-7}+ i\,3.15 \times 10^{-17}$  MeV${}^{1/2}$  remains negligible compared to the internal counterpart, which does not depend on the ANCs.

 \begin{table}[htb]
\caption{Channel and internal radiaitive width amplitudes for the SR  $1^{-}.$    }
\begin{center}
\begin{tabular}{|c|c|c|c|}
\hline
  $E$  & $\gamma_{\gamma\, 0}^{1\,(ch)}(E)$ & $\gamma_{\gamma\, 0}^{1\,(m)}\,(int)$  & $\gamma_{\gamma\, 0}^{1\,(p)}\,(int)$    \\ 
(MeV)     &     (MeV${}^{1/2}$)     &  (MeV${}^{1/2}$) &  (MeV${}^{1/2}$)\\
\hline
 $0.1$   &   $3.20 \times 10^{-8} + i\,7.16 \times 10^{-30}$ &  $-0.00146$   &    $0.00146$              \\
\hline
$0.2$ & $3.22 \times 10^{-8}   + i\,1.2 \times 10^{-21} $  &    $ -0.00146$     & $
0.00146
$        \\
\hline
$0.3$  & $3.25 \times 10^{-8}  +  i\,4.97 \times 10^{-18}$ &   $-0.00146$   &    $0.00146$     \\
\hline
$0.4$ & $3.28 \times 10^{-8}  +  i\,6.75\times 10^{-16}$ &   $-0.00146$   &    $0.00146$   \\
\hline
$0.5$ & $3.32 \times 10^{-8}  +  i\,1.85 \times 10^{-14}$ &  $-0.00146$   &    $0.00146$    \\
\hline
\end{tabular}
\end{center}
\label{RDwa1}
\end{table} 

\subsubsection{Subthreshold resonance  $2^{+}$  }
\label{SR2}

Although for $L=2$  the effective charge is not small, due to the high binding energy of the ground state of ${}^{16}{\rm O},\,$    $\,\gamma_{\gamma\, 0}^{2}\,(ch)$  is significantly smaller than
$\gamma_{\gamma\, 0}^{2}\,(int)$, see Table \ref{RDwa2}.  Hence  Eq.  (\ref{GgE1}) is also applicable for  $L=2$. 
The experimental value  of the radiative width of the subthreshold bound state  $2^{+}$
decaying to the ground state  is  $\Gamma_{\gamma\, 0}^{2}(-0.245\,{\rm MeV})= (9.7 \pm 0.3)\times 10^{-8}$  MeV  \cite{TUNL}. 
Again, as it was for the $1^{-}$  subthreshold state,  Eqs. (\ref{GgE1}) and (\ref{ggint2}) are used to calculate the internal radiative width amplitudes (notations are the same as in Table \ref{RDwa1}).  
A $3\%$ 
uncertainty of $\Gamma_{\gamma\, 0}^{2}(-0.245\,{\rm MeV})$
propagates to $1.5\%$ uncertainty  of  $\gamma_{\gamma\, 0}^{2}\,(int)$. For calculations  we use
$\gamma_{\gamma\, 0}^{2\,(p)}\,(int)  = 0.0089 \pm 0.0001$  MeV${}^{1/2}$.  
It should be underscored that  for the high subthreshold ANC for the $2^{+}$ state  (the last row of Table \ref{Table_ANCs})  and  the high ground-state ANC of $337$ fm${}^{-1/2}$ \cite{Shen}  the channel radiative width amplitude  $\gamma_{\gamma\, 0}^{2\,(ch)}(300\,{\rm keV})= 1.63 \times10^{-3}+ i\,6.67 \times 10^{-14}$  MeV${}^{1/2}$  remains small but not negligible compared to the internal counterpart.

 \begin{table}[htb] 
\caption{Channel and internal radiative width amplitudes for the SR  $2^{+}.$  }
\begin{center}
\begin{tabular}{|c|c|c|c|}
\hline
  $E$  & $\gamma_{\gamma\, 0}^{2\,(ch)}(E)$ & $\gamma_{\gamma\, 0}^{2\, (m)}\,(int)$  & $\gamma_{\gamma\, 0}^{1\,(p)}\,(int)$    \\ 
(MeV)     &     (MeV${}^{1/2}$)     &  (MeV${}^{1/2}$) &  (MeV${}^{1/2}$)\\
\hline
 $0.1$   &   $1.5 
 \times
 10^{-5} +i\,8.13 \times 10^{-28}$ &  $- 0.0089$   &    $ 0.0089$              \\
\hline
$0.2$ & $1.5 \times 10^{-5} +i\,1.43 \times 10^{-19}$  &    $- 0.0089$    &   $ 0.0089$        \\
\hline
$0.3$  & $1.5 \times 10^{-5} +i\,6.19 \times 10^{-16}$ &   $- 0.0089$   &    $ 0.0089$     \\
\hline
$0.4$ & $1.5 \times 10^{-5} +i\,8.81 \times 10^{-14}$ &   $- 0.0089$   &      $ 0.0089$   \\
\hline 
$0.5$ & $1.5 \times 10^{-5} +i\,2.53 \times 10^{-12}$ &   $- 0.0089$  &   $ 0.0089$    \\
\hline 
\end{tabular}
\end{center}
\label{RDwa2}
\end{table} 

\section{$S$-factors for   ${}^{12}{\rm C}(\alpha, \,\gamma){}^{16}{\rm O}$ radiative capture  to the ground state of ${}^{16}{\rm O}$}
\label{Sfctrs1}

\subsection{$S$-factors for resonance  $E1$ and $E2$  captures to the ground state}
\label{SE1E2}

 First we present the results of the calculations for two dominant low-energy  $S$-factors corresponding to the  resonance $E1$ and $E2$ transitions to the ground state.  
 The resonance $E1$  transition  is  made of contributions from the SR  $(7.12\, {\rm MeV}, 1^{-}),\,$  
 the first above the threshold 
 resonance   $(9.59\, {\rm MeV}, 1^{-}),\,$ and higher  $1^{-}$ resonances.
 Similarly, the  resonance  $E2$  transition  is dominated by the SR $(6.92\, {\rm MeV}, 2^{+})$, 
 the lowest above-threshold   
 resonance $(9.85\, {\rm MeV}, 2^{+})$ and higher $2^{+}$  resonances.  

Although below for convenience we show the $S$-factors for the radiative capture ${}^{12}{\rm C}(\alpha, \,\gamma){}^{16}{\rm O}$ for energies $E < 3$ MeV,  our main focus is the low-energy region  $E < 1$ MeV, and especially the most effective astrophysical energy $\,E=300$ keV,  at which we can check the impact of the subthreshold  ANCs  on the  $S(300\, {\rm keV})$-factors.  We compare the $S$-factors calculated using  the central values of the ANCs in the last row of  Table  \ref{Table_ANCs} with the results from the review \cite{deBoer}.   
In this energy region it is enough to take into account  the two lowest levels 
(the SR and the lowest resonance at $E>0$)
in addition to background resonances for the $E1$ and $E2$  resonance transitions.
 Details and parameters used in the calculations  are given in \cite{deBoer}.
 
\begin{figure}[hb]
\includegraphics[width=\columnwidth]{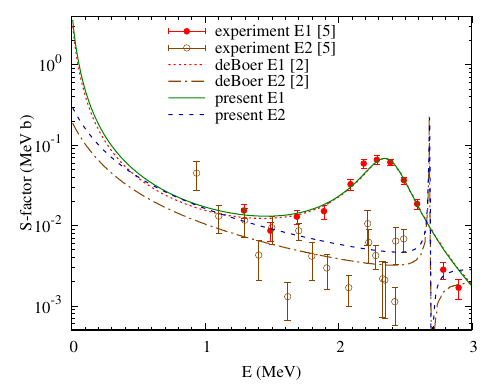}
\caption{ The $S$-factors for the   ${}^{12}{\rm C}(\alpha, \,\gamma){}^{16}{\rm O}$ radiative capture reaction to the ground state  through the $1^{-}$ and $2^{+}$ resonances. }
\label{fig_SfctrE1E2} 
\end{figure}

 Thus the resonance $E1$ and $E2$ $\,S$-factors depicted in Fig. \ref{fig_SfctrE1E2}  
  are calculated
  for the subthreshold ANCs  $2.08 \times 10^{14}$ fm${}^{-1/2}$  (for the $(7.12\, {\rm MeV}, 1^{-})\,$)  and  $\,1.14 \times 10^{5}$  fm${}^{-1/2}$  (for the $(6.92\, {\rm MeV}, 2^{+})\,$)  
  from \cite{deBoer},  which are compared with  the higher sets of the ANCs, $2.27 \times 10^{14}$  fm${}^{-1/2}$  and 
$1.42 \times 10^{5}$  fm${}^{-1/2}$ 
\cite{Blokh2023}   (the last row of Table  \ref{Table_ANCs}).
It can be seen that as  the values  of the subthreshold  ANCs increase, both $E1$ and $E2$ $S$-factors exhibit an increase.  
The exact values  of the calculated resonance $E1$ and $E2$ $S$-factors at $300$ keV   are given in Table \ref{Table_results}. 

We need to add  some additional comments about the  calculation of  the $E1$ and $E2$ resonance captures to the ground state of ${}^{16}{\rm O}$. 
Precisely speaking, we accounted for the interference of the $E1$ and $E2$ resonance captures to the ground state with the  $E1$ and $E2$ direct captures to the ground state.
Shown in  Fig.  \ref{fig_SfctrE1E2}  $\,E1$ and $E2$  
$S$-factors  include all the transitions to the ground state  except for  the cascade transitions, which are 
discussed in subsection \ref{Cascade1}.
 In the  calculations shown in Fig. \ref{fig_SfctrE1E2}   we adopted  a low value of the ground-state ANC $C_{0}= 58$ fm${}^{-1/2}$ taken from \cite{deBoer},  because our goal was to compare with the results from \cite{deBoer} by varying only the subthreshold ANCs.  
 
\subsection{Radiative capture to subthreshold states and the total $S$-factor  for 
${}^{12}{\rm C}(\alpha, \,\gamma){}^{16}{\rm O}$  radiative capture}
\label{Cascade1}

Besides the $E1$ and $E2$  resonance captures to the ground state, we also calculated  the cascade transitions to the ground state through  the four excited bound states  $(6.05\, {\rm MeV},\,0_{2}^{+})$,  $\,(6.13\, {\rm MeV},\, 3^{-}),\,$   $(6.92\, {\rm MeV},\,2^{+})\,$  and $(7.12\, {\rm MeV},\,1^{-})$   using the ANCs from the last two rows of Table  \ref{Table_ANCs}.  The cascade transitions to the ground state of ${}^{16}{\rm O}$ represent two-step processes: 
the direct capture to one of the excited states  is followed by the radiative decay of  the excited bound state to the ground state.  The ANCs of the four subthreshold bound states under consideration govern  the normalization of the direct capture amplitudes.
 
\begin{figure}[h]
\includegraphics[width=\columnwidth]{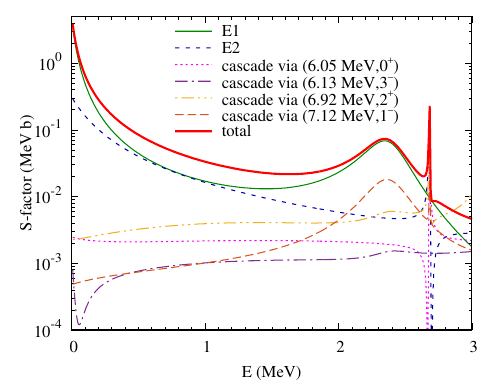}
\caption{ All the calculated $S$-factors for radiative captures to the ground state.
The total $S$-factor given by the sum of all the $S$-factors  shown in this figure.}
\label{tot_Sfctrs}
\end{figure}

Figure  \ref{tot_Sfctrs} depicts all the $S$-factors calculated  using the  ANCs obtained by the extrapolation procedure.
In this figure we show not only the  $S$-factors  for the cascade transition through  the  four excited states of ${}^{16}{\rm O}$ but also the $S$ factors for the $E1$ and $E2$ resonance captures to the ground state of  ${}^{16}{\rm O}$ (see Fig. \ref{fig_SfctrE1E2})  and the total $S$-factor,  which is the sum of  all six $S$-factors 
shown
in Fig. \ref{tot_Sfctrs}.

\section{Correlation between the subthreshold  ANCs and the total $E1$ and $E2$ $S$-factors}
\label{Results1}

The results of the comparison of the $S$-factors from the current paper and the ones from \cite{deBoer} are presented in Table  \ref{Table_results}. The difference between 
them
is caused by variation of only one parameter, the ANC of the corresponding subthreshold state. 
It is evident that the rise in  subthreshold state ANCs leads to an increase in the $E1$ and $E2$ $S(300\,{\rm keV})$-factors.

\begin{table}[htb]
\caption{Comparison of the current $S$-factors and the $S$-factors from \cite{deBoer} for transition to the ground state of ${}^{16}{\rm O}$ at the most effective astrophysical energy of $300$ keV.  
The resonance $E1$ and $E2$  transitions include interference with the direct $E1$ and $E2$ captures 
to  the ground state. The cascade transition is the radiative capture to the ground state proceeding through four subthreshold bound states of ${}^{16}{\rm O}$: $0_{2}^{+},\,3^{-},\,2^{+}$ and  $\,1^{-} .$   We remind that all the calculations were done for the low ground-state ANC of $58$ fm${}^{-1/2}.$
The S-factors are given in units of keVb.  }
\begin{center}
\begin{tabular}{|c|c|c|}
\hline
 Transition to the ground state via 
 & $S(300\, {\rm keV})$  & $S(300\, {\rm keV})$ \\ 
\hline 
   resonance  
   +  direct capture 
                   &   Present    &  Ref. \cite{deBoer}          \\
                    \hline
$E1$   &   $98$   &    85  \\
$E2$ & $70$  & $45$        \\
$E1$ + $E2$ & $168$  & $130$           \\
cascade   &   Present    &  Ref. \cite{deBoer}          \\
  $0_{2}^{+}+3^{-} +
2^{+}  +1^{-} $             &    $6$     &    $7$                 \\
total    &   Present    &  Ref. \cite{deBoer}    \\
     $E1+E2$ +cascade         &   $174$    &       $137$           \\
\hline
\end{tabular}
\end{center} 
\label{Table_results}
\end{table}

 \begin{table}[htb]
\caption{Correlation between  the uncertainties of squares of ANCs and the $S(300\, {\rm keV})$-factors.}
\begin{center}
\begin{tabular}{|c|c|c|}
\hline
  Transitions & $\Delta C_{l}^{2}$,  $\%$ & $\Delta S(300\, {\rm keV})$, $\%$  \\ 
\hline 
 $E1$   &   $19$   &  15   \\
$E2$ & $55$  & $56$        \\
$E1$ + $E2$ &  & $29$           \\
\hline   
\end{tabular}
\end{center}
\label{Table_uncertainties}
\end{table} 

Table \ref{Table_uncertainties}  shows the correlation between the variations of the ANCs of the subthreshold bound states causing the corresponding  variations $S(300\, {\rm keV})$-factors. 
In this  table $\Delta C_{l}^{2} = C_{l (pr)}^{2}/  C_{l (RMP) }^{2} -1$   and   $\Delta S(300\, {\rm keV}) = S_{pr}(300\, {\rm keV})/ S_{RMP}(300\, {\rm keV}) -1.$ Quantities  with the subscript "pr" are calculated using the present  ANCs  and those with the subscript "RMP" use ANCs from \cite{deBoer}. 
We observe around $79\%$ correlation between the uncertainties of the  ANC for the  $1^{-}$ subthreshold state and of  the total  $S_{E1}(300\, {\rm keV})$-factor for the $E1$ transition to the ground state of ${}^{16}{\rm O}$.  An additional contribution to the $E1$ transition may come  from  interference between  the   $1^{-}$  SR with the higher broad  $\,(E_{X}= 9.59\,{\rm MeV}; 1^{-})$ MeV  resonance and the direct capture  to the ground  state.\footnote{The impact of the  interference of the SR  with the  direct capture  will be checked below by using  the higher ground-state ANC of $337$ fm${}^{-1/2}.$}  Thus, a   $5\%$  uncertainty of the subtheshold ANC $C_{1}$   generates  about $4\%$  uncertainty of the total  $E1$  $S(300\, {\rm keV})$-factor.  
As previously discussed, the $E1$ $S_{E1}(300\,{\rm keV})$-factor is independent of the ground-state ANC, as illustrated in Table \ref{Table_ANCSR} below.  Hence the interference  of the $E1$ SR  occurs only with the broad $9.59$ MeV resonance.

For the $E2$ transition  we get  almost $100\%$  correlation between the subthreshold  ANC for the subthreshold $2^{+}$ state  and the   $S_{E2}(300\, {\rm keV})$-factor  for  the $E2$  transition  to the ground state.  We can conclude that at low  ground-state ANC of $58$ fm${}^{-1/2}$, the interference effect of the subthreshold   $2^{+}$ resonance with 
higher resonance and
the direct capture amplitudes is negligible.
Hence, when other parameters are fixed,  the uncertainty  of the subthreshold ANC  for the  $2^{+}$  state almost completely  propagates  into the uncertainty of the  $S_{E2}(300\, {\rm keV})$  for the  $E2$ radiative  captures  to the ground state. Accordingly, for the low ground-state ANC,  a $5\%$ uncertainty of the subthreshold ANC $C_{2}$  generates  about  $5\%$ uncertainty in the  total   $\,S_{E2}(300\, {\rm keV})$-factor. 

 \section{Low-energy $S$-factors for  resonance capture to the ground state through subthreshold  
 resonances }
\label{Subrescontr1}
\subsection{Contribution of the subthreshold  $\,1^{-}\,$ and $\,2^{+}\,$ resonances}

While  the importance of subthreshold ANCs in the analysis  of the  $\,{}^{12}{\rm C}(\alpha, \,\gamma){}^{16}{\rm O}$  reaction was acknowledged  and  large amount of the literature on this subject has been published (see \cite{deBoer}  and references therein),  the precise contribution of the SRs had not been determined previously.
  From Table  \ref{Table_uncertainties}    we can already draw some preliminary conclusions about contribution of the pure SRs into the total $S(300\,{\rm keV})$-factor.  Below  we  present the calculations of  the low-energy  $S$-factors   for the $E1$ and $E2$ transitions through the subthreshold  $\,1^{-}\,$ and $\,2^{+}\,$ resonances  for capture to the  ground state. No other mechanisms are included. It allows us to evaluate numerically the contribution of the SRs  to the total  low-energy $S$-factor, and, in particular,  $S(300\,{\rm keV})$-factors. We then can determine the role of the subthreshold ANCs and show the necessity to decrease their uncertainties to  improve the uncertainty of the total  $S(300\,{\rm keV})$-factor.  

\begin{figure}[tbp]
\includegraphics[width=1.0\columnwidth]{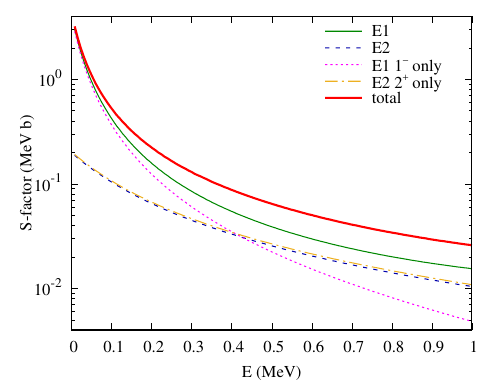}
\caption{The low-energy  $S$-factors for the   ${}^{12}{\rm C}(\alpha, \,\gamma){}^{16}{\rm O}$ reaction calculated  using all the subthreshold  and ground-state  ANCs  from \cite{deBoer} (see Table \ref{Table_ANCs}).  
}
\label{fig_SsubdeBoer} 
\end{figure}

\begin{figure}[tbp]
\includegraphics[width=1.0\columnwidth]{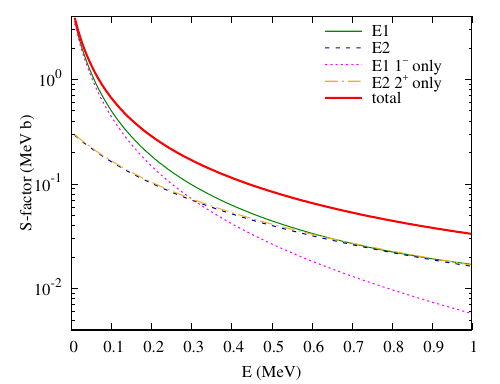}
\caption{The low-energy  $S$-factors for the   ${}^{12}{\rm C}(\alpha, \,\gamma){}^{16}{\rm O}$ reaction calculated  using
the  ground-state  ANC  $C_{0}(0.0\,{\rm MeV}) = 58$ fm${}^{-1/2}\,$  \cite{deBoer}, while all other ANCs   are  taken from  the last two rows 
of Table  \ref{Table_ANCs}.  
}
\label{fig_SsubBlokh} 
\end{figure}

\begin{figure}[tbp]
\includegraphics[width=1.0\columnwidth]{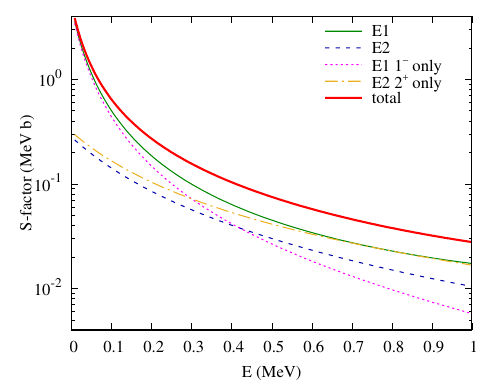}
\caption{The low-energy  $S$-factors for the   ${}^{12}{\rm C}(\alpha, \,\gamma){}^{16}{\rm O}$ reaction calculated for the  ground-state ANC
 $C_{0}(0.0\,{\rm MeV}) = 337$ fm${}^{-1/2}\,$  \cite{Shen}.    
 All other the ANCs  are  taken from  the last two rows  of Table  \ref{Table_ANCs}.}
\label{fig_SsubBlokhShen} 
\end{figure}

Figures \ref{fig_SsubdeBoer}-\ref{fig_SsubBlokhShen}   depict the results of the  calculations.
In Fig. \ref{fig_SsubdeBoer}    we use the low ANCs from   \cite{deBoer}  (see Table  \ref{Table_ANCs})  including the 
 ground-state ANC  $C_{0} = 58$ fm${}^{-1/2}.\,$     
In Fig.  \ref{fig_SsubBlokh}   the  ground-state  ANC  remains  low (the same  as in Fig. \ref{fig_SsubdeBoer}), while  the subthreshold  ANCs   are higher than in Fig. \ref{fig_SsubdeBoer}  (the last row  of Table  \ref{Table_ANCs}). In Fig. \ref{fig_SsubBlokhShen}    we use higher subthreshold  ANCs  (as in Fig.  \ref{fig_SsubBlokh}) and higher  the ground-state ANC, which is  taken from \cite{Shen}:    $C_{0} = 337$ fm${}^{-1/2}.\,$  

One can see that the $E1$ and $E2$  subthreshold $1^{-}$  and $2^{+}$ resonance  transitions to the ground state  of ${}^{16}{\rm O}$  give the dominant  contribution  to the total low-energy $S$-factor confirming   a pivotal role of the  subthreshold   ANCs  in the calculation of the low-energy  $S$-factor.  

The $S$-factors  depicted in Figs.  \ref{fig_SsubdeBoer} and \ref{fig_SsubBlokh}  show very similar behavior near the energy of $300$ keV:  the   $S$-factors $\,S_{E1}^{(SR)}(E)$ for the $E1$ transition  to the ground state   proceeding through the   $1^{-}\,$  SR  are slightly lower than the total $E1$ $\,S$-factor $\,S_{E1}(E).\,$ 
It agrees with our findings  about correlation between the ANC of the subthreshold state $1^{-}$ and the total  $\,S_{E1}(E)$-factor  presented  in  Table \ref{Table_uncertainties}. However, according to Figs. \ref{fig_SsubdeBoer} and \ref{fig_SsubBlokh}, as well as Table \ref{Table_ANCSR}, it is evident that the value of $\,S_{E1}^{(SR)}(300\,{\rm keV})$ rises as the subthreshold ANC for the $1^{-}$ state increases.

The $S_{E2}^{(SR)}(300\,{\rm keV})$-factor  for the radiative capture  through the SR $2^{+}$  exceeds the total  $S_{E2}(300\, {\rm keV})$-factor  only by $3\%.$   It means that  we observe a small destructive interference  of the  amplitude for the $E2$  transition to the ground state through the SR $2^{+}$ 
with the higher broad $2^{+}$ resonances and  the direct capture amplitudes.  

Figure  \ref{fig_SsubBlokhShen} displays the results obtained using  the  higher  subthreshold ANCs (the last two rows in Table \ref{Table_ANCs})   and the  higher  ground-state  ANC of $337$ fm${}^{-1/2}$ \cite{Shen}.  
One can see that  the behavior of  the  $S_{E1}^{(SR)}(E)$-factor  is  very similar to  the one from  previous figures  proving that this  $S$-factor practically  does not depend on the  ground-state ANC.  The independence of the   $S_{E1}(E)$-factor on the ground-state ANC  follows from a simple fact that this dependence can come only  from the channel  radiative transition amplitude $M_{SR}^{(ch)}(E)$   of the SR $1^{-}$ and direct capture amplitude. However, both contain almost vanishing effective charge, which suppresses the contribution of   both amplitudes.

The same is true for the SR  $S_{E2}^{(SR)}(300\,{\rm keV})$-factor  for the $E2$ transition, which is not sensitive to the ground-state ANC.  Still, the total $S_{E2}(300\,{\rm keV})$-factor shows a noticeable decrease as the ground-state ANC increases. It means that for the $E2$ transition  we observe  the destructive  interference  of the capture through the SR with the direct transition, which depends on the ground-state ANC.  For the small  ground-state ANC of $58$ fm${}^{-1/2}$ this interference  is  very small but  it becomes significant for the higher ground-state ANC.  

The numerical results of the calculations presented in Figs.  \ref{fig_SsubdeBoer}-\ref{fig_SsubBlokhShen} are summarized in   Table  \ref{Table_ANCSR}.  
We see  that  increase  of the subthreshold  ANCs  increases  $\,S_{E1}^{(SR)}(300\,{\rm keV})$  and
$S_{E2}^{(SR)}(300\,{\rm keV})$. 
 However,  the contribution of  the  $\,S_{E1}^{(SR)}(300\,{\rm keV})$  ($7$th column) to  the   budget of the total   $S_{E1}(300\,{\rm keV})$  changes very little,   $71-74\%$  and   does not depend on  the ground-state ANC. 
 
 The $S_{E2}^{(SR)}(300\,{\rm keV})$ also  does  not depend on the ground-state ANC because the channel radiative width amplitude for the $E2$ transition is very small.  But the  total $\,S_{E2}(300\,{\rm keV})\,$  depends on the ground-state  ANC due to the  destructive interference of the  $E2\,$  SR amplitude with the direct capture amplitude to the ground state of ${}^{16}{\rm O}$. It results in  decrease of the  $\,S_{E2}(300\,{\rm keV})\,$  compared  to  the  $S_{E2}^{(SR)}(300\,{\rm keV})$  ($8$-th column in Table \ref{Table_ANCSR}).
 
 \begin{table}[htb] 
\caption{The $S(300\,{\rm keV})$-factors (expressed in keVb)  for the radiative capture ${}^{12}{\rm C}(\alpha, \,\gamma){}^{16}{\rm O}$   presented in Figs. \ref{fig_SsubdeBoer}, \ref{fig_SsubBlokh} and  \ref{fig_SsubBlokhShen}.   The first column pertains to the figure displaying the relevant results. $\,S_{E1}\,$ and $\,S_{E2}$  are the total $S$-factors for the $E1\,$ and $E2$  transitions,  $S_{E1+E2}= S_{E1} + S_{E2}$.  $S_{E1}^{(SR)}$ and $S_{E2}^{(SR)}$   are the  $S(300\,{\rm keV})$-factors  for the resonance $E1$ and $E2$  transitions  through the SRs $1^{-}$  and $2^{+}$ only.   $S_{E1}^{(SR)}/S_{E1}$  and    $S_{E2}^{(SR)}/S_{E2}$ are the ratios of the corresponding $S$-factors. }
\begin{center}
\begin{tabular}{|c|c|c|c|c|c|c|c|}  
\hline
  Fig. & $S_{E1+E2}$  &  $S_{E1}$ & $S_{E2}$ &    $S_{E1}^{(SR)}$  &  $S_{E2}^{(SR)}$  &  $S_{E1}^{(SR)}/S_{E1}$  &  $S_{E2}^{(SR)}/S_{E2}$   \\ 
\hline 
  \ref{fig_SsubdeBoer}  &  $130$   &   $85$   &  $45$   &  $60$  &   $46$  &  $0.71$     &  $1.02$    \\
 \ref{fig_SsubBlokh}    &  $168$    & $98$  & $70$   &   $72$   &   $ 72$     &   $ 0.74 $      &    $1.03$          \\
 \ref{fig_SsubBlokhShen}  &  $155$ &   $99$   &  $56$      &   $72$    &   $72$     &  $0.73$     &    $1.29$         \\
\hline   
\end{tabular}
\end{center}
\label{Table_ANCSR}
\end{table}

\subsection{Impact of the ground-state ANC}

Here, we revisit the role of the ground-state ANC once again.
 In this paper  we  employed  two   ground-state ANCs,  $58$ fm${}^{-1/2}$  \cite{deBoer}  and  $337$  fm${}^{-1/2}$  \cite{Shen}.  
 
As we mentioned above, for the low  ground-state ANC  the interference of the resonance $E1$ and $E2$  transitions  with the direct captures to the ground state  is small.  The present calculations  of the  
 $E1$ and $E2$ transitions  with the  ground-state ANC of $337$  fm${}^{-1/2}$   and  the subthreshold  $1^{-}$ and $2^{+}$  ANCs from  \cite{Blokh2023}  (see also Table \ref{Table_ANCs}), which include the  interference of the  resonances  and  direct captures,  resulted   in  $S_{E1}(300\,{\rm keV}) = 99\,$ keVb and   $S_{E2}(300\,{\rm keV}) = 56\,$  keVb. 
 
Our findings indicate that  a notable rise  in the ground-state ANC  of   ${}^{16}{\rm O}$  has a minimal impact  on 
the astrophysical $S_{E1}(300\,{\rm keV})$-factor  (compare the last two rows in column $S_{E1}$ of Table  \ref{Table_ANCSR}). However, it does lead to a $20\%$  reduction  of  the  $S_{E2}(300\,{\rm keV})$  compared with  the $S_{E2}(300\,{\rm keV})$  obtained for the lower ground-state  ANC  (compare  the last two rows in column  $S_{E2}$ of Table  \ref{Table_ANCSR}).  

Meanwhile, the uptick in the subthreshold ANCs boosts both $S_{E1}(300\,{\rm keV})$ and $S_{E2}(300\,{\rm keV})$ values.  Therefore, we can infer that the concurrent increase of the subthreshold and ground-state ANCs leads to a boost of the $S_{E1}(300\,{\rm keV})$ and $S_{E2}(300\,{\rm keV})$ values  mitigated by a reduction of the latter at higher ground-state ANC.  
From Table \ref{Table_ANCSR}  we can see that  the total $S$-factor  $S(300\,{\rm keV})= 130$ keVb calculated for low subthreshold  and the ground-state ANCs elevates to  $168$ keVb for higher subthreshold ANCs and low the ground-state ANC, but it decreases to $155$ keVb for the higher subthreshold and ground-state ANCs.

Hence our calculations confirm a correlation between the ground-state ANC of ${}^{16}{\rm O}$ and the ANC of the subthreshold $2^{+}$ state, in agreement with \cite{Shen}.  From this perspective, taking into account the results from \cite{deBoer} and adopting the ANC for the $2^{+}$ state from \cite{Blokh2023},  we can assume  that the ground-state ANC of ${}^{16}{\rm O}$, which fits the S -factor  from \cite{deBoer}, should be less than 337 fm${}^{-1/2}$. 

Finally,  we would like to point out that one of the main uncertainties  of the ground-state ANC  from \cite{Shen}, which we use in this paper,  stems from the uncertainty of the ANC for ${}^{11}{\rm B} \to \alpha+  {}^{7}{\rm Li}$, which was not  properly discussed in \cite{Shen}.  That is why  at this stage, due to the lack of  reliable ANC for the ground state of ${}^{16}{\rm O}$,  we cannot provide the total uncertainty of the ground-state ANC  of  ${}^{16}{\rm O}$ extracted in \cite{Shen}. 

\begin{figure}[htb]
\includegraphics[width=1.0\columnwidth]{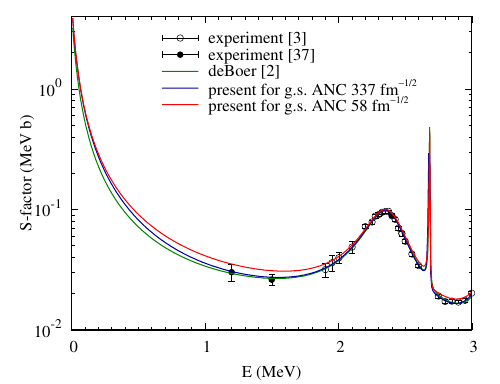}
\caption{The total $S$-factors for the   ${}^{12}{\rm C}(\alpha, \,\gamma){}^{16}{\rm O}$ reaction given by $E1+E2 + {\rm cascade}$ transitions to the ground state of  ${}^{16}{\rm O}$  for three different calculations. 
The experimental data are from~\cite{Kettner} and \cite{Yama}.}
\label{fig_Sfctrtotal1} 
\end{figure}

Figure  \ref{fig_Sfctrtotal1}  shows the total $S$-factors contributed by the sum of the $E1+E2$ resonance transitions to the ground state of ${}^{16}{\rm O}$  (including the interference with the direct captures)  plus the cascade transition to the ground state  of ${}^{16}{\rm O}$. 
There are many  experimental  datasets available in the literature \cite{deBoer}. We show one of them as a representative  example. 
The explanation of the three different  lines is given in the caption to the figure.
It is evident that  increasing the ground-state ANC and the subthreshold ANCs is bringing the present results closer to those in \cite{deBoer}, which were obtained using lower ANCs.

\section{Summary}
\label{Summary1}
The $S$-factors for the  ${}^{12}{\rm C}(\alpha, \,\gamma){}^{16}{\rm O}$  reaction are calculated within the $R$-matrix AZURE2 code  using the recently determined ANCs for four subthreshold states $0_{2}^{+},\,3^{-},\,2^{+},\,1^{-}$ \cite{Blokh2023,BKMS5}, which represent the high end of the ANCs, and a low ground-state ANC of $58$ fm${}^{-1/2}$  of   ${}^{16}{\rm O}$ \cite{deBoer}.  The results are compared with the those  obtained in  \cite{deBoer}  using  the subthreshold ANCs  available in the literature, which represents the lower end of the ANCs.  
Our comprehensive calculations of the low-energy $S$-factors encompass the $E1$ and $E2$ SR transitions to  the ground state of $^{16}$O, which interfere with the  higher resonances and direct captures,  and cascade radiative captures to the ground state of $^{16}$O through four subthreshold states: $0_2^+,\,3^-,\, 2^+$ and $1^-$.  Since our ANCs are higher than those used by 
deBoer {\it et al.} 
\cite{deBoer},
the present total  $S$-factor at the most effective astrophysical energy of 300 keV  is  174 keVb  versus  137 keVb of that work.
Higher subthreshold bound-state ANCs used in the present calculations lead to a higher $S(300\,  {\rm keV})$-factor and higher low-temperature reaction rates (see Appendix \ref{Appendix2}). 

Since in the present calculations, all the parameters, except for the subthreshold ANCs, were taken from  \cite{deBoer},  we are able to check  the dependence  of the $S$-factors for  the  ${}^{12}{\rm C}(\alpha, \,\gamma){}^{16}{\rm O}$  radiative capture at very low energies, where  the resonance captures through the subthreshold state dominate.
The contributions of the  $E1$ and $E2$ SR transitions  to the total $S$-factors  $S_{E1}(300\,{\rm keV})$  ($S_{E2}(300\,{\rm keV})$)  are  $\,71\%$ and $74\%$  ($102\%$ and  $103\%$)  for the  subthreshold  ANCs from \cite{deBoer} and  present  ones, respectively.  
Moreover, we also observe that  for the uncertainty  of the $1^{-}$  ($2^{+}$)  subthreshold ANC of $19\%$   ($55\%$)  generates $15\%$ ($56\%$) uncertainty  in the  $S_{E1}(300\,  {\rm keV})$ ($S_{E2}(300\,{\rm keV})$).  It is essential to emphasize  that the ratio $S_{E1}^{(SR)}(300\,{\rm keV})/S_{E1}(300\,  {\rm keV}) =74\%$  is  less than the  correlation of the uncertainties $\Delta C_{1}^{2}$  and    $\Delta S_{E1}(300\, {\rm keV})$, which is $79\%$.   This arises from  the additional  contribution of the interference  of the  $E1$  SR and  the broad  ($9.59\,{\rm MeV}, 1^{-}$)  resonance, which also depends on $C_{1}$. 

 The  $E1$ transition of the SR  $1^{-}$ is not influenced by the ground-state ANC as the $E1$  nonresonant capture is the isospin forbidden transition  (with a negligibly small effective charge), yet it still interferes constructively with a broad   $(9.585\,{\rm MeV};\,1^{-})$  resonance  giving (for the present subthreshold ANCs)  an additional  $26\%$ contribution to the total  $S_{E1}(300\,{\rm keV})$-factor.  
 
 The interference between the $E2$ transition of the SR $2^{+}$ and direct capture is minimal when the ground-state ANC is small, but becomes destructive at higher ground-state ANC, resulting in a contribution of $-29\%$.
The low-energy $S_{E2}(300\,{\rm keV})$-factor experiences a smaller increase when both subthfreshold and the ground-state ANCs rise together due to their anticorrelation, compared to when only the subthreshold ANCs increase. 
 
 In summary, our key findings indicate that:\\
 (1)  For given parameters of the broad  $1^{-}$ resonance  $9.59$ MeV \cite{deBoer}  we can control the uncertainty  of the    $\,S_{E1}(300\,{\rm keV})$-factor, which stems from the uncertainty of the square of the subthreshold ANC  $C_{1}^{2}$,  as the total uncertainty  of the former is  $79\%$   of the latter. \\
 (2)  The uncertainty in the ground-state ANC limits the conclusiveness of our analysis for the $E2$ transition compared to the $E1$ case.  The most pressing issue that needs attention is determining the ground-state ANC.

 Nevertheless, we can draw some preliminary conclusions.
For the low ground-state  ANC, the uncertainty of the  $\,S_{E2}(300\,{\rm keV})$-factor  is  completely generated by the  uncertainty  of $C_{2}^{2}$.   Hence  the total uncertainty  of the $S(300\,{\rm keV})$-factor is given by   $\sqrt{ [0.79\, {\Delta}C_{1}^{2}]^{2}  +  [{\Delta}C_{2}^{2}]^{2}  }.\,$   An increase in the ground-state ANC results in a decrease in the 
contribution of the $\,S_{E2}(300\,{\rm keV})$  to the total $S(300\,{\rm keV})$-factor, making the uncertainty in $C_{2}^{2}$ less important.  Therefore, we can assert that the upper bound of the total uncertainty for the $S(300\, \rm{keV})$ is $\sqrt{ [0.79\, {\Delta}C_{1}^{2}]^{2} + [{\Delta}C_{2}^{2}]^{2} }$.

\begin{acknowledgments}  
R.J.D. utilized resources from the Notre Dame Center for Research Computing and was supported by the National Science Foundation through grant
Nos. PHY-1713857 and PHY-2011890, and the Joint Institute for Nuclear Astrophysics through grant No. PHY-1430152
(JINA Center for the Evolution of the Elements).
A.S.K. acknowledges the support from the Australian Research Council. \\
\end{acknowledgments}

\appendix
\section{ Reduced width of the subthreshold resonance}
\label{Appendix1} 
Discussing the appearance of the Whittaker function in Eq. (\ref{gSR1}) is appropriate and ties in with deriving the expression for the reduced width of the SR.
In order to establish a relationship between $\Gamma_{J_{s}}^{SR}(E),\,\,\gamma_{l_{s}}^{SR}(R_{ch})$ and the ANC $C_{l_{s}}$ of the subthreshold bound state, one can extrapolate the elastic scattering $S$-matrix (see Eq. (\ref{Smtrpole1}))  to the SR pole.

The fundamental concept of the $S$-matrix's analyticity requires the existence of a bound-state pole positioned on the upper part of the imaginary axis in the $k$-plane. In the vicinity of the subthreshold pole in the $k$-plane, the $S$-matrix for elastic scattering is represented by  \cite{mukblokh2022}
 \begin{align}
 {\mathbb S}_{l_{s}}  \stackrel{k \to i\,\kappa_{s} +0}{=} - i(-1)^{l_{s}}\,e^{i\,\pi\,\eta_{s}}\,\frac{ C_{l_{s}}^{2} }{k- i\,\kappa_{s}}.
 \label{explSmtrpole1}
 \end{align} 
In this equation, the pole's residue is articulated in terms of the subthreshold ANC $C_{l_{s}}$. 

At the same time, the elastic scattering $S$-matrix in the $R$-matrix approach can be written in terms of the resonance width or, alternatively, in terms of the reduced width. In order to make the derivation of the $S$-matrix in the presence of the SR  clearer, we start from the beginning.  In the $R$-matrix formalism,  the single-level, single-channel $S$-matrix is provided by
\begin{align}
 {\mathbb S}_{l_{s}}  = e^{{-2\,i\,(\delta_{c\,l_{s}}^{hs} - \sigma_{cl_{s}}^{C})}}\,\frac{1/R_{l_{s}}  - [\Delta_{l_{s}}(E)- B_{l_{s}} - i\,P_{l_{s}}(k,\,R_{ch})] }{1/R_{l_{s}}  - [\Delta_{l_{s}}(E)- B_{l_{s}} + i\,P_{l_{s}}(k,\,R_{ch})},  
  \label{SmslschRm1}
 \end{align} 
 where  $R_{l_{s}}= {\tilde \gamma}^{2}_{l_{s} }/(E_{s}- E)$ is the $R$-matrix,  ${\tilde \gamma}^{2}_{l_{s} }$ is the formal  reduced width of the subthreshold level.  Additionally,  $\Delta_{l_{s}}(E)$ is the Thomas shift  and   $B_{l_{s}}$ is the boundary condition.  We remind that $J_{s}=l_{s}$.
 $E_{s}$ is the $R$-matrix level energy  corresponding to the subthreshold resonance. We adopt  $E_{s}=-\varepsilon_{s}$ and $B_{l_{s}}= \Delta_{l_{s}}(-\varepsilon_{s})$. Then for $E$ in the vicinity of $-\varepsilon_{s}$
 \begin{align}
 {\mathbb S}_{l_{s}}  \approx e^{{-2\,i\,(\delta_{c\,l_{s}}^{hs} - \sigma_{cl_{s}}^{C})}}\, \frac{-\varepsilon_{s} -E  + i\,P_{l_{s}}(k,\,R_{ch})\, \gamma^{2}_{l,_{s}}}{-\varepsilon_{s} -E  - i\,P_{l_{s}}(k,\,R_{ch})\, \gamma^{2}_{l_{s}}}.  
  \label{SmslschRm2}
 \end{align} 
We employed  $\Delta_{l_{s}}(E)- \Delta_{l_{s}}(-\varepsilon_{s}) \approx  \frac{{\rm d}\Delta_{l_{s}}(E)}{{\rm d}E}\big|_{E=-\varepsilon_{s}}\,(E +\varepsilon_{s})$. It allows one to introduce the observable reduced width  
\begin{align}
 \gamma^{2}_{l_{s}}=  \frac{{\tilde \gamma}^{2}_{l_{s} }}{1+ \frac{{\rm d}\Delta_{l_{s}}(E)}{{\rm d}E}\big|_{E=-\varepsilon_{s}}  }.
 \label{obsgmsq1}
 \end{align}
 
 When extrapolating  $E \to -\varepsilon_{s}$  Eq. (\ref{SmslschRm2})  is reduced to
\begin{align}
 {\mathbb S}_{l_{s}}   \stackrel{E \to -\varepsilon_{s}}{\approx} - 2\,i\,e^{{-2\,i\,(\delta_{c\,l_{s}}^{hs} - \sigma_{cl_{s}}^{C})}}\,\frac{P_{l_{s}}(k, R_{ch})\,[\gamma_{l_{s}}^{SR}(R_{ch})]^{2}}{ E   + \varepsilon_{s}  }. 
\label{SmSRlimit1}
 \end{align} 
 To obtain it we  took into account that  in the denominator at $E \leq 0$  $\,\Gamma_{J_{s}}^{SR}(E)=0$   due to the presence in it  the barrier penetrability $P_{l}(k,\,R_{ch})$ (see Eq. (\ref{GammaSR1})), which represents the imaginary component of the logarithmic derivative of the wave function that remains real for negative energies.  However, taking into account Eq. (\ref{deltahs1}), the numerator is made up of 
\begin{align}
 e^{2\,i\,(\sigma_{l_{s}}^{C} -  \delta_{l_{s}}^{hs}  )}\,P_{l_{s}}(k, R_{ch})= & \frac{k\,R_{ch}}{[O_{l_{s}}(k,\,R_{ch})]^{2}}    \nonumber\\
=& \frac{(-1)^{l_{s}}\,e^{-\pi\,\eta\,{\rm sign\,Re}k}\,k\,R_{ch}}{ \big[W_{-i\,\eta, l_{s}+1/2}(-2\,i\,k\,R_{ch})\big]^{2} }.
 \label{Plextr1}
 \end{align}
 $e^{\pi\,\eta\,{\rm sign\,Re}k}\,[W_{-i\,\eta, l_{s}+1/2}(k,\,R_{ch})\big]^{2} $  is analytic function in  the upper half $k$ plane (${\rm Im}k >0$) except the point $k=0\,$  \cite{mukblokh2022}.    Its extrapolation to the subthreshold bound state pole is simple, leading to the emergence of the Whittaker function $W_{-\eta_{s},\,l_{s}+1/2}(2\,\kappa_{s}\,R_{ch})$ for the subthreshold bound state:
\begin{align}
  e^{2\,i\,(\sigma_{l_{s}}^{C} -  \delta_{l_{s}}^{hs}  )}\,P_{l_{s}}(k, R_{ch}) 
  \stackrel{k \to i\,\kappa_{s}}{=} \frac{(-1)^{l_{s}}\,e^{i\,\pi\,\eta_{s}}\,i\,\kappa_{s}\,R_{ch}}{ \big[W_{-\,\eta_{s}, l_{s}+1/2}(2\,\kappa_{s}\,R_{ch})\big]^{2} }.
 \label{Plextr1}
 \end{align}
With Eq. (\ref{Plextr1}) considered,  in the vicinity of the subthreshold bound state pole in the $k$ plane Eq. (\ref{SmSRlimit1}) transforms to
\begin{align}
& {\mathbb S}_{l_{s}}   \stackrel{k \to i\,\kappa_{s}}{\approx}  -2\,i\,\mu\,R_{ch}\,\frac{(-1)^{l_{s}}\,e^{i\,\pi\,\eta_{s}}\,}{ \big[ W_{-\eta_{s}, l_{s}+1/2}\big]^{2} }   
 \frac{[\gamma_{l_{s}}^{SR}(R_{ch})]^{2}}{k- i\,\kappa_{s}}. 
\label{Smslsch2}
 \end{align} 
Equation (\ref{gSR1}) is derived from the comparison of Eqs. (\ref{explSmtrpole1}) and (\ref{Smslsch2}).

\section{ Reaction Rates}
\label{Appendix2}

In Table  \ref{Table_ReactionRates}  we compare the reaction rates  calculated using the present total $S$-factor with the ones from \cite{deBoer}.  For easier comparison, the tabulated low-temperature reaction rates are calculated at the same temperatures as in  \cite{deBoer}.
Since the present  $S$-factor is  larger than that from \cite{deBoer}, at low temperatures $T_{9}  <2.0 $  our reaction rates  exceed  the reaction rates from \cite{deBoer}.  However, since we constrain our calculations to low-energy $S$-factor,  at temperatures  $T_{9} \geq 2$   the reaction rates from \cite{deBoer} exceed ours. 

 \begin{table}[htb]
\caption{Low temperature reaction rates  ($T_{9}  \leq 2$)}
\begin{center}
\begin{tabular}{|c|c|c|} 
\hline
  $T_{9}$ & Reaction Rates   ($\frac{ {\rm cm}^{3}}{{\rm s}\,{\rm mol}}$)  &  Reaction Rates  ($\frac{ {\rm cm}^{3}}{{\rm s}\,{\rm mol}}$)\\
   &       Present   &  Ref. \cite{deBoer}  \\
\hline 
$0.06$	&   $6.81 \times  10^{-26}$    &    $6.78 \times  10^{-26}$        \\
$0.07$      &  $3.65 \times 10^{-24}$      &   $3.28 \times 10^{-24}$       \\
$0.08$      &  $9.33 \times 10^{-23}$      &   $8.00 \times 10^{-23}$       \\
$0.09$      &  $1.42 \times 10^{-21}$      &   $1.18 \times 10^{-21}$       \\
$0.1$      &  $1.46 \times 10^{-20}$      &   $1.20 \times 10^{-21}$       \\
 $0.11$   &   $1.11 \times 10^{-19}$    &     $9.03 \times 10^{-20}$           \\
 $0.12$   &   $6.66 \times 10^{-19}$    &     $5.38 \times 10^{-19}$           \\
 $0.13$   &   $3.29 \times 10^{-18}$    &     $2.65 \times 10^{-18}$           \\
  $0.14$   &   $1.39 \times 10^{-17}$    &     $1.11 \times 10^{-17}$           \\
  $0.15$   &   $5.11 \times 10^{-17}$    &     $4.08 \times 10^{-17}$           \\
  $0.16$   &   $1.68 \times 10^{-16}$    &     $1.34 \times 10^{-16}$           \\
  $0.18$   &   $1.37 \times 10^{-15}$    &     $1.09 \times 10^{-16}$           \\
 $0.20$   &   $8.34 \times 10^{-15}$     &     $6.64 \times 10^{-15}$ \\
$0.30$ & $4.68 \times10^{-12}$       &   $3.73 \times 10^{-12}$   \\
$0.35$ & $4.10 \times10^{-11}$       &   $3.28 \times 10^{-11}$   \\
 $0.4$   &   $2.44 \times 10^{-10}$    &  $1.96 \times 10^{-10}$ \\
 $0.45$   &   $1.10 \times 10^{-9}$    &  $8.82 \times 10^{-10}$ \\
$0.5$ & $3.99 \times 10^{-9}$    &      $3.22 \times 10^{-9}$      \\
$0.6$ & $3.33 \times 10^{-8}$    &      $2.70 \times 10^{-8}$      \\
$0.7$ & $1.80 \times 10^{-7}$    &      $1.47 \times 10^{-7}$      \\
$0.8$ & $7.2 \times 10^{-7}$    &      $5.92 \times 10^{-7}$      \\
$0.9$ & $2.32 \times 10^{-6}$    &      $1.92 \times 10^{-6}$      \\
$1.0$ & $6.37 \times 10^{-6}$    &      $5.30 \times 10^{-6}$      \\
$1.25$ & $4.85 \times 10^{-5}$    &      $4.10 \times 10^{-6}$      \\
$2.0$ & $2.60 \times 10^{-3}$    &      $2.40 \times 10^{-3}$      \\
\hline
\end{tabular}
\end{center}
\label{Table_ReactionRates}
\end{table} 



\begin{thebibliography}{99}
\bibitem{RolfsRodney}  C. E. Rolfs  and W. S. Sydney, Cauldrons in the Cosmos,  The University of Chicago Press, 1988.
\bibitem{deBoer}   R. J. deBoer, J. Görres, and M. Wiescher,
R. E. Azuma, A. Best, C. R. Brune, C. E. Fields, S. Jones, M. Pignatari, 
D. Sayre, K. Smith, F. X. Timmes, Rev. Mod. Phys. {\bf 89},   035007   (2017).
\bibitem{Kettner} D. Schürmann, A. Di Leva, et al., Eur. Phys. J. A  \textbf{26}, 301 (2005).
\bibitem{redder} A. Redder, H.W. Becker, C. Rolfs, H.P. Trautwetter,  T.R. Donoghue, T. C. Rinckel,  J. W. Hammer  and K.  Langanke,  Nucl. Phys. A{\bf 462},    385    (1987).
\bibitem{Kremer} R. M. Kremer, C. A. Barnes, K. H. Chang, H. C. Evans, B. W. Filippone, K. H. Hahn, and L. W. Mitchell, Phys. Rev. Lett. {\bf 60},  1475   (1988).
\bibitem{BarkerKajino} F. C. Barker and T. Kajino, Austr. J. Phys. {\bf 44},,   369   (1991).
\bibitem{Ouellet}   J. M. L. Ouellet, H. C. Evans, H. W. Lee, J. R. Leslie, J. D. MacArthur, W. McLatchie, H.-B. Mak, P. Skensved, J. L. Whitton, X. Zhao, and T. K. Alexander, Phys. Rev. Lett. {\bf 69},  1896   (1992). 
\bibitem{azuma} R. E. Azuma, L. Buchmann, F. C. Barker, C. A. Barnes, J. M. D$'$Auria, M. Dombsky, U. Giesen, K. P. Jackson, J. D. King, R. G. Korteling, P. McNeely, J. Powell, G. Roy, J. Vincent, T. R. Wang, S. S. M. Wong, and P. R. Wrean
Phys. Rev. C {\bf 50},   1194  (1994).
 \bibitem{Brune}  C. R. Brune, W. H. Geist, R. W. Kavanagh, and K. D.Veal, Phys. Rev. Lett. \textbf{83},  4025    (1999). 
\bibitem{kunz}  R. Kunz, M. Jaeger, A. Mayer, J. W. Hammer, G. Staudt, S. Harissopulos, and T. Paradellis, Phys. Rev. Lett. {\bf 86},    3244   (2001).
\bibitem{assuncao} M. Assun\c c\~ao, M. Fey, A. Lefebvre-Schuhl, J. Kiener, V. Tatischeff, J. W. Hammer, C. Beck, C. Boukari-Pelissie, A. Coc, J. J. Correia, S. Courtin, F. Fleurot, E. Galanopoulos, C. Grama, F. Haas, F.  Hammache, F.Hannach, S. Harissopulos,  A. Korichi, R.Kunz, D. LeDu, A. Lopez-Martens, D. Malcherek, R. Meunier, Th. Paradellis, M. Rousseau, N. Rowley, G. Staudt, S. Szilner,  J. P. Thibaud, and J. L. Weil, Phys. Rev.  C {\bf 73}  055801    (2006).
\bibitem{Balhout}  A. Belhout, S. Ouichaoui,  H. Beaumevieille, A. Boughrara, S. Fortier, J. Kiener, J. M. Maison, S. K. Mehdi, L.Rosier, J. P. Thibaud,   A.Trabelsi, J. Vernotte, Nucl. Phys. A {\bf 793},   178   (2007).
\bibitem{plug} R. Plag,  R. Reifarth, M. Heil, F. Käppeler, G. Rupp, F. Voss, and K. Wisshak,
 Phys. Rev.  C {\bf 86},    015805   (2012).
\bibitem{sayre} D. B. Sayre, C. R. Brune, D. E. Carter, D. K. Jacobs, T. N. Massey, and J. E. O'Donnell, Phys. Rev. Lett. {\bf 109},   142501
(2012).
\bibitem{Schurmann}  D. Sch\"{u}rmann,  L.  Gialanella,  R. Kunz, F. Striederd,   Phys. Lett. {\bf B711},   35  (2012).
\bibitem{gai2013} M. Gai, Phys. Rev. {\bf C 88},     06280(R)  (2013).
\bibitem{Avila} M. L. Avila, G. V. Rogachev, E. Koshchiy, L. T. Baby, J. Belarge, K. W. Kemper, A. N. Kuchera, A. M. Mukhamedzhanov, D. Santiago-Gonzalez, and E. Uberseder,  Phys. Rev. Lett. {\bf 114},   071101   (2015).
\bibitem{Shen} Y. Shen, B. Guo, R. J. deBoer, E. Li, Zh.  Li, Y.   Li, X. Tang, D. Pang,
S.  Adhikari, Ch. Basu, J. Su, Sh. Yan, Q. Fan, J.  Liu, Ch.  Chen, Zh.  Han,
X. Li, G. Lian, T. Ma, W. Nan, Weike Nan, Y.  Wang, Sh. Zeng, H. Zhang, and W. Liu, ApJ  {\bf 945},    41   (2023).
\bibitem{Blokh2017}   L. D. Blokhintsev, A. S. Kadyrov, A. M. Mukhamedzhanov, and D. A. Savin,   Phys. Rev. C {\bf 95} , 044618   (2017).
\bibitem{Blokh2018}   L. D. Blokhintsev, A. S. Kadyrov, A. M. Mukhamedzhanov, and D. A. Savin,   Phys. Rev. C {\bf 97},    024602   (2018).
\bibitem{BKMS5}    L. D. Blokhintsev, A. S. Kadyrov, A. M. Mukhamedzhanov, and D. A. Savin,  Eur. Phys. J. A  {\bf  58},  257   (2022).
\bibitem{Blokh2023}   L. D. Blokhintsev, A. S. Kadyrov, A. M. Mukhamedzhanov, and D. A. Savin,  Eur. Phys. J. A  {\bf  59},   162     (2023).
\bibitem{Dubovichenko}  S. B. Dubovichenko, A. S. Tkachenko, R. Ya. Kezerashvili, N. A. Burkova, and A. V. Dzhazairov-Kakhramanov,  Phys. Rev. C {\bf 105}, 065806 (2022). 
\bibitem{muk2023}   A. M. Mukhamedzhanov, Eur. Phys. J. A  {\bf 59},  43    (2023).
\bibitem{mukblokh2022}   A. M. Mukhamedzhanov and L. D. Blokhintsev,  Eur. Phys. J. A  {\bf 58},  29   (2022). 
\bibitem{muktribble1999}  A. M. Mukhamedzhanov and R. E. Tribble, Phys. Rev. C {\bf 59},  3418   (1999).
\bibitem{Oulevsir} N. Oulevsir,   F. Hammache, P. Roussel, M. G. Pellegriti, L. Audouin, D. Beaumel, A. Bouda, P. Descouvemont, S. Fortier, L. Gaudefroy, J. Kiener, A. Lefebvre-Schuhl, and V. Tatischeff, Phys. Rev. C \textbf{85},  035804  (2012) .
\bibitem{Orlov1}  
Yu. V. Orlov, B. F. Irgaziev, and L. I. Nikitina, Phys. Rev. C \textbf{93},  014612   (2016).
\bibitem{Orlov2} 
Yu. V. Orlov, B. F. Irgaziev, and Jameel-Un Nabi, Phys. Rev. C \textbf{96}, 025809  (2017).
\bibitem{Sparen}
O. L. Ram\'{\i}rez Su\'arez and J.-M. Sparenberg, Phys. Rev. C {\bf 96},   034601   (2017).
\bibitem{Ando} 
Shung-Ichi Ando, Phys. Rev. C \textbf{97},   014604   (2018) .
\bibitem{Orlov3}   
Yu. V. Orlov, Nucl. Phys. A \textbf{1014},  122257   (2021);   Erratum, Nucl. Phys. A \textbf{1018},  122385  (2022).
\bibitem{Hebborn} 
C. Hebborn, M. L. Avila, K. Kravvaris, G. Potel, and S. Quaglioni, arXiv:2307.05636v2 (2023).
\bibitem{TUNL}  
D. R. Tilley,  H. R. Weller, and C. M. Cheves, Nucl. Phys. A {\bf 564}, 1  (1993).
\bibitem{AisenbergGreiner}  J. M. Eisenberg  and W. Greiner,  Excitation Mechanisms Of The Nucleus. Electromagnetic and Weak Interactions,  North-Holland Publishing Company, Amsterdam-London, 1970.
\bibitem{Baye}  D. Baye and E. Tursunov,  J. Phys. G {\bf 45}, 085102 (2018),
\bibitem{Yama} H. Yamaguchi, K. Sagara, K. Fujita, D. Kodama, Y. Narikiyo, K. Hamamoto, T. Ban, N. Tao and T. Teranishi, AIP Conference Proceedings \textbf{1594}, 229 (2014).


\end{thebibliography}
\end{document}